\documentclass[a4paper,11pt]{article}

\usepackage[a4paper]{geometry}
\usepackage{color}
\usepackage{graphicx}
\usepackage{amsmath}
\usepackage{amsfonts}
\usepackage{amsthm}
\usepackage{amscd}
\usepackage[mathcal]{euscript}
\usepackage{epsfig}
\usepackage{amssymb}
\usepackage{slashed}
\usepackage{tabularx}
\usepackage{calligra}
\usepackage{enumerate}
\usepackage{hyperref}
\usepackage{float}
\usepackage{stackrel}
\usepackage[T1]{fontenc}
\usepackage{longtable}
\usepackage{amsthm}
\usepackage{mathrsfs}
\usepackage{verbatim} % for comments
\usepackage{fancyhdr}
\usepackage{tikz}
\usepackage{tikz-cd}
\usepackage{slashed}
\usetikzlibrary{matrix}
\usetikzlibrary{positioning}
\usepackage[all]{xy}
\usepackage{units}
\usepackage{textcomp}
\usepackage{turnstile}
\usepackage{vmargin}
\usepackage{anysize}
\usepackage{ stmaryrd }

\usepackage{tikz-3dplot}

\usepackage[font=small]{caption}

\numberwithin{equation}{section}

\setmargins{2,5cm}       % left margin
{1,5cm}                        %  top margin
{16cm}                      %  text widht
{23,42cm}                    % text height
{10pt}                           % header height
{1cm}                           % space between text and header
{0pt}                             % footer height
{2cm}                           %  space between text and footer

\theoremstyle{plain}
\newtheorem{theorem}{Theorem}[section]
\newtheorem{proposition}[theorem]{Proposition}

\newtheorem{remark}[theorem]{Remark}

\newcommand{\tn}{\tilde{\nabla}}
\newcommand{\bn}{\bar{\nabla}}
\newcommand{\tT}{\tilde{\Theta}}
\newcommand{\T}{\Theta}

\DeclareMathOperator{\tho}{\text{\rm\th}}
\DeclareMathOperator{\edt}{\text{\rm\dh}}

\begin{document}

\title{Parallel spinors, pp-waves, and gravitational perturbations}

\author{Bernardo Araneda\footnote{Email: \texttt{bernardo.araneda@aei.mpg.de}}  \\
Max-Planck-Institut f\"ur Gravitationsphysik \\ 
(Albert Einstein Institut), Am M\"uhlenberg 1, \\
D-14476 Potsdam, Germany}

\date{\today}

\maketitle

\begin{abstract}
We prove that any real, vacuum gravitational perturbation of a 
4-dimensional vacuum pp-wave 
space-time can be locally expressed, modulo gauge 
transformations, as the real part of a Hertz/Debye potential, 
where the scalar potential satisfies the wave equation. 
We discuss relations with complex perturbations, 
complex space-times, non-linear structures, and real 
spaces with split (ultra-hyperbolic/Kleinian) signature.
Motivated by generalized notions of parallel spinors,
we also discuss generalizations of the result to other space-times.
\end{abstract}

\section{Introduction}

Pp-wave space-times are exact solutions to the Einstein equations 
modelling gravitational radiation. 
These space-times are interesting both physically and 
mathematically for many reasons: 
they are relevant for gravitational wave physics; 
they satisfy, in appropriate cases, a linear superposition principle;
they represent a universal limit for general relativity 
in that, as shown by Penrose \cite{Penrose}, 
every Lorentzian space-time looks like a pp- 
(plane\footnote{Plane waves are a special case of pp-waves: 
the former have an isometry group that is at least 5-dimensional, 
while the latter possess in general only one Killing vector.}) 
wave near a null geodesic; all their curvature invariants vanish 
(which is relevant e.g. for string theory); etc. 
In addition, closer to our motivation, they represent the 
simplest case of a 4-dimensional Lorentzian geometry that 
admits a parallel spinor field \cite{Bryant}. 
In this work we study vacuum gravitational perturbations of 
pp-waves in four dimensions, and
the problem of representations of 
solutions to linearized gravity in terms of the so-called 
Hertz/Debye potentials.

\smallskip
The general question we want to address is: 
can any real vacuum gravitational perturbation 
be represented, modulo gauge, in terms of a 
Hertz/Debye potential?
This is conjectured to be true, locally, for perturbations of 
all algebraically special vacuum spaces, cf. the introduction 
in \cite{AAAW}; but, as far as we know, 
the problem has only been completely solved for the 
case of Minkowski space-time \cite{TorresdelCastillo}, 
\cite{Penrose65}, \cite[Section 5.7]{PR1}. 
Our main result is given in sections \ref{Sec:MainResult} 
and \ref{Sec:halfkahler}.

\smallskip
Although in this work we study the special case of pp-waves, 
the techniques we use also apply to perturbations of the 
above more general class of solutions. 
This is because our procedure is based on exploiting the 
existence of special geometric structures called 
$\alpha$- and $\beta$-surfaces, or simply {\em twistor 
surfaces}, that are present, in particular, for any algebraically 
special Einstein space-time. 
Pp-waves have the advantage that, while having
a very simple curvature structure that facilitates computations,
the conceptual difficulties one has to deal with in the other 
more complicated cases are already present in this class. 
We illustrate this point by studying the more general case of 
a ``half-K\"ahler'' vacuum space-time (see sections 
\ref{Sec:Motivation} and \ref{Sec:halfkahler}).

\smallskip
Furthermore, the fact that our method is based on twistor surfaces 
allows us to give, in the pp-wave case, a precise description 
of the close connection that exists between the 
Hertz/Debye representation of real linear gravitational 
perturbations and the fully non-linear geometry of 
a complex analogue of a pp-wave: 
a complex 4-manifold admitting a parallel spinor field. 
Notably, the situation can also be understood in terms of 
{\em real} geometry, but for a metric with split 
(also called ultra-hyperbolic, Kleinian, or neutral) signature.

\smallskip
Finally, parallel spinors constitute the major motivation in this work, 
since as detailed in section \ref{Sec:Motivation} below, 
a pp-wave is the simplest 
case of a general scheme in which special geometries 
(including e.g. black hole space-times)
are characterized in terms of ``generalized parallel spinors''. 
Our approach exploits a simple link between 
generalized parallel spinors and complex geometry, 
and it has direct connections 
to the twistor programme and the heavenly formalisms 
of Penrose, Newman and Pleba\'nski; 
see section \ref{Sec:ParallelSpinors}.

\medskip 
\noindent
{\bf Summary.} 
In section \ref{Sec:ParallelSpinors} we give an elementary 
review of spinors in 4d; present our motivation 
relating parallel spinors and complex geometry; 
and deduce the structure of a 4-geometry that admits 
a parallel spinor in Lorentz signature and also for complex metrics.
Our main result is presented in section \ref{Sec:Perturbations} 
where we study gravitational perturbations. 
In section \ref{Sec:halfkahler} we study a generalization of this 
result, to the case of a ``half-K\"ahler'' vacuum space-time.
Some final remarks are given in section \ref{Sec:Conclusions}. 
We include appendix \ref{Appendix:Gauge} with additional 
details of calculations.
We follow the notation and conventions of 
Penrose and Rindler \cite{PR1, PR2}; in particular, we 
use abstract indices.

\section{Parallel spinors, real and complex space-times}
\label{Sec:ParallelSpinors}

\subsection{Preliminaries}
\label{Sec:Preliminaries}

Given a 4d complex vector space with 
a metric $g_{ab}$ and an orientation, the orthogonal group is
\begin{align}
 {\rm SO}(4,\mathbb{C}) = 
 ({\rm SL}(2,\mathbb{C})\times{\rm SL}(2,\mathbb{C}))
 /\mathbb{Z}_{2}.
 \label{spingroup}
\end{align}
The relation between the two sides of \eqref{spingroup} is 
understood 
by fixing an isomorphism $\sigma$ between $\mathbb{C}^{4}$ 
and $\mathbb{C}^{2}\otimes\mathbb{C}^{2}$, i.e. one 
writes a column vector $v$ in $\mathbb{C}^{4}$ as 
a matrix $\sigma(v)$ in $\mathbb{C}^{2}\otimes\mathbb{C}^{2}$. 
Then \eqref{spingroup} means that, for any orthogonal 
transformation $\Lambda\in{\rm SO}(4,\mathbb{C})$, there 
are elements $L$ and $R$ in ${\rm SL}(2,\mathbb{C})$ such 
that $\sigma(\Lambda v)=L\sigma(v)R^{\rm t}$.

Elements in each copy of $\mathbb{C}^{2}$ are called spinors.
Since each $\mathbb{C}^{2}$ has an independent action of 
${\rm SL}(2,\mathbb{C})$, there are two different kinds of spinors. 
We say that the two kinds have opposite `chirality'.
In abstract indices, these are 
distinguished by primed and unprimed indices, e.g.
$\psi^{A'}$ and $\varphi^{A}$, and the isomorphism $\sigma$ 
is $v^{a} \to \sigma(v)^{AA'}\equiv v^{AA'}$. 
We usually omit $\sigma$, so that we identify 
$v^{a}\equiv v^{AA'}$. 
This way we have the usual identification of indices 
$a=AA'$, $b=BB'$, etc., which we follow in this work.
From the relation $g(v,v)=2\det\sigma(v)$ one deduces that 
the metric is $g_{ab}=\epsilon_{AB}\epsilon_{A'B'}$, 
where $\epsilon_{AB}$ is the natural volume element of 
$\mathbb{C}^{2}$.

Without any reality conditions, spinors of opposite chirality 
are independent.
Real forms of \eqref{spingroup} corresponding 
to different metric signatures are recovered by using 
different reality structures. These structures can in turn 
be understood as operations on spinors, that we call `spinor 
conjugations', and they may or may not lead to relations between 
chiralities.

\smallskip
For Lorentzian reality conditions, spinor conjugation 
interchanges chirality, 
so the action of the two factors in the RHS of \eqref{spingroup} 
is not independent, and one recovers the Lorentz group 
$\rm{SO}(1,3)={\rm SL}(2,\mathbb{C})/\mathbb{Z}_{2}$.
We denote Lorentzian spinor conjugation with an overbar, e.g. 
$\varphi^{A}\to\bar{\varphi}^{A'}$.
A spinor $\varphi^{A}$ and its complex conjugate 
$\bar{\varphi}^{A'}$ produce a real null vector 
$N^{a}=\varphi^{A}\bar{\varphi}^{A'}$.
Given a basis of $\mathbb{C}^{2}$, $\{o^{A},\iota^{A}\}$, 
one can consider the complex conjugate basis 
$\{\bar{o}^{A'},\bar{\iota}^{A'}\}$ and construct  
four linearly independent null vectors as 
\begin{align}
 \ell^{a}=o^{A}\bar{o}^{A'}, \qquad n^{a}=\iota^{A}\bar{\iota}^{A'}, \qquad m^{a}=o^{A}\bar{\iota}^{A'}, \qquad \bar{m}^{a}=\iota^{A}\bar{o}^{A'}.
 \label{nulltetrad}
\end{align}
If the basis $\{o^{A},\iota^{A}\}$ is normalized by 
$\epsilon_{AB}o^{A}\iota^{B}=1$, then the vectors \eqref{nulltetrad} 
satisfy the usual conditions for a null tetrad:
$g_{ab}\ell^{a}n^{b}=1=-g_{ab}m^{a}\bar{m}^{b}$,
and the rest vanishes.

\smallskip
For Euclidean (/Riemannian) 
reality conditions, spinor conjugation $\dagger$ 
preserves chirality, but a spinor $\varphi^{A}$ 
and its complex conjugate $\varphi^{\dagger A}$
are linearly independent: if $\varphi^{A}$ has components 
$(a,b)$ relative to some basis, then 
$\varphi^{\dagger A}$ has components $(-\bar{b},\bar{a})$. 
Since $\dagger$ is anti-linear and it holds $\dagger^{2}=-1$, 
this is really a quaternionic structure. 
The Euclidean form of \eqref{spingroup} is 
${\rm SO}(4,\mathbb{R})=
({\rm SU}(2)\times{\rm SU}(2))/\mathbb{Z}_{2}$, 
and chiralities are independent.
Given a spinor $o^{A}$, one has a spin basis 
$\{o^{A},o^{\dagger A}\}$, but unlike Lorentz signature, 
this does not give a basis for the opposite chirality.

\smallskip
Finally, the restriction to real elements in ${\rm SL}(2,\mathbb{C})$ 
corresponds to a metric with split signature. The isomorphism 
\eqref{spingroup} becomes 
${\rm SO}(2,2)=({\rm SL}(2,\mathbb{R})
\times{\rm SL}(2,\mathbb{R}))/\mathbb{Z}_{2}$, 
spinors are real and chiralities are independent.

\smallskip
Over an open neighbourhood on a smooth manifold $M$ 
equipped with a metric $g_{ab}$, 
one constructs the primed and unprimed spinor bundles 
$\mathbb{S}'$, $\mathbb{S}$, and 
the considerations above apply pointwise on each fiber. 
A spinor field is a (local) section of $\mathbb{S}$ or 
$\mathbb{S}'$ (or tensor products of them). 
If $x^{a}$ are local coordinates on $M$, 
we use the identification of indices $a=AA'$, etc. to 
write e.g. ${\rm d}x^{a}\equiv{\rm d}x^{AA'}$, 
so the metric is $g=\epsilon_{AB}\epsilon_{A'B'}
{\rm d}x^{AA'}\otimes{\rm d}x^{BB'}$.
Similarly, the Levi-Civita connection is $\nabla_{a}=\nabla_{AA'}$.
If $(M,g_{ab})$ is real, the operator $\nabla_{AA'}$ is also real.

\subsection{Motivation: parallel spinors}
\label{Sec:Motivation}

A parallel (or covariantly constant) spinor is a spinor field 
$o^{A}$ that satisfies
\begin{equation}
 \nabla_{AA'}o^{B} = 0. \label{PS}
\end{equation}
The existence of a non-trivial solution to \eqref{PS} imposes 
strong restrictions on the geometry. 
Specific restrictions depend on the metric signature, see 
\cite{Bryant}. 

In Lorentz signature, complex conjugation of \eqref{PS} 
gives a parallel spinor with opposite chirality, 
$\nabla_{AA'}\bar{o}^{B'} = 0$.
The real null vector $\ell^{b}=o^{B}\bar{o}^{B'}$ is 
therefore covariantly constant, so the geometry is a pp-wave.
In this work we are interested in this case, see 
section \ref{Sec:ppwaves}.
In Riemann signature, the complex conjugate of \eqref{PS} is 
$\nabla_{AA'}o^{\dagger B}=0$. 
One then has a parallel spin frame, so the manifold must be 
hyper-K\"ahler. We will not focus on this case. 
In split signature, a (real) solution to \eqref{PS} is equivalent 
to a null K\"ahler structure, see \cite{Dunajski, DP}.

\medskip
Our interest in parallel spinors actually arises from ``generalized'' 
versions of them, where one considers connections more 
general than the Levi-Civita connection.
Such generalizations are important both in physics and in 
mathematics.
For example, these objects appear in supergravity 
in relation to the existence of supersymmetries;
and they are also relevant in certain areas of 
pure geometry, for instance concerning different definitions 
of `mass'. See e.g. \cite{Tod} \footnote{We are 
interested in parallel {\em Weyl} spinors, while in supergravity 
and related areas one considers {\em Dirac} spinors.}.

But our major motivation is the connection that 
generalized parallel spinors turn out to have with complex 
geometry.
For example, a K\"ahler manifold can be characterized 
by the existence of a parallel (pure) {\em projective} spinor, 
cf. \cite{Lawson}. 
In four dimensions (where all spinors are pure), this can be 
expressed in terms of the Riemannian version of a connection 
that is well-known in general relativity, the so-called 
`GHP' connection $\Theta_{AA'}$. 
Interestingly enough, a Hermitian manifold can be similarly 
defined via parallel spinors, using a generalization of 
$\Theta_{AA'}$, that we may call 
`complex-conformal connection' or `conformally invariant 
GHP connection', and we denote by $\mathcal{C}_{AA'}$, 
cf. \cite{Araneda1, Araneda2}.
We summarize the situation in table \ref{table}.

\begin{table}
\centering
\begin{tabular}{|c|c|c|c|}
\hline
Condition & Riemann signature & Lorentz signature & 
Split signature \\ \hline 
$\nabla_{AA'}o^{B}=0$ & hyper-K\"ahler & pp-wave & 
null-K\"ahler \\ 
$\Theta_{AA'}o^{B}=0$ & K\"ahler & ``half-K\"ahler'' & 
no name \\
$\mathcal{C}_{AA'}o^{B} = 0$ & Hermitian & ``half-Hermitian'' & 
no name \\
\hline
\end{tabular}
\caption{Different notions of parallel spinors give different
special 4-geometries. 
$\Theta_{AA'}$ is the `GHP connection', 
and $\mathcal{C}_{AA'}$ is a conformally invariant version of it. 
Apart from $\nabla_{AA'}o^{B}=0$, the other equations are 
non-linear, since the connections depend on $o^{A}$.
The terminology in the Lorentzian case is perhaps 
not standard, although similar names have been used  
by Flaherty \cite{Flaherty}. 
In the split case, the conditions can be related
to {\em para-complex} geometry.}
\label{table}
\end{table}

The operators $\Theta_{AA'}$ and $\mathcal{C}_{AA'}$ 
are well-defined in any signature\footnote{One needs a 
{\em pair} of spinors 
in the construction of $\Theta_{AA'}$, $\mathcal{C}_{AA'}$. 
In the Riemannian case a single spinor is enough since its 
complex conjugate gives the other. 
In the other cases the extra spinor can be chosen at will.}.
Our interest in the (generalized) parallel spinor equations 
presented in table \ref{table} is that they imply the existence of 
twistor surfaces, which are the basic object that 
give integration procedures.
The kind of algebraic and differential manipulations that 
one has to follow in these procedures is essentially 
the same in all cases, which is why we find 
the parallel spinors viewpoint attractive: it is both 
conceptually (geometrically) meaningful and 
computationally practical.
In this paper we are interested in the simplest case, 
eq. \eqref{PS}, and its applications to the linearized gravity 
problem in general relativity. 
We will also discuss the ``half-K\"ahler'' case, 
see section \ref{Sec:halfkahler}.
For the treatment of $\mathcal{C}_{AA'}o^{B}=0$ in 
 conformal geometry, see \cite{Araneda2} (perturbations are 
 not treated in this reference).

\subsection{Lorentz signature: pp-wave space-times}
\label{Sec:ppwaves}

We define a pp-wave space-time as a 4-dimensional 
Lorentzian manifold $(M,g_{ab})$ that admits a non-trivial 
parallel real null vector $N^{a}$, $N_{a}N^{a}=0$, 
$\nabla_{a}N^{b}=0$, 
and such that the Ricci tensor is $R_{ab}\propto N_{a}N_{b}$.
As shown in \cite{Aichelburg}, any such geometry admits 
a parallel spinor, that in the rest of this work 
we denote by $o^{A}$. The associated parallel 
null vector is denoted by $\ell^{a}=o^{A}\bar{o}^{A'}$.

\smallskip
The following result is just the standard characterization of 
pp-waves in terms of Brinkmann coordinates, and it is 
well-known:
\begin{proposition}\label{Prop:ppwave}
Let $(M,g_{ab})$ be a Lorentzian space-time admitting a 
non-trivial parallel spinor field $o^{A}$, eq. \eqref{PS}. 
Then there exist (locally) a coordinate system 
$(u,v,\zeta,\bar{\zeta})$ and a real scalar field 
$H=H(v,\zeta,\bar\zeta)$ such that the metric is 
\begin{equation}
 g = 2({\rm d}u{\rm d}v-{\rm d}\zeta{\rm d}\bar{\zeta}) 
 + H{\rm d}v^{2}.
 \label{ppwavemetric}
\end{equation}
The Ricci scalar vanishes, and the rest of the curvature is 
given by
\begin{align}
 \Phi_{ABA'B'}={}& \tfrac{1}{2}H_{\zeta\bar{\zeta}} 
  o_{A}o_{B}\bar{o}_{A'}\bar{o}_{B'}, 
  \label{Riccippwave}\\
 \Psi_{ABCD} ={}& \tfrac{1}{2}H_{\bar{\zeta}\bar{\zeta}} 
  o_{A}o_{B}o_{C}o_{D}.
  \label{Weylppwave}
\end{align}
\end{proposition}

It is instructive to look at the derivation of this result from the 
perspective of twistor surfaces; we will do this in 
the rest of this subsection.

\smallskip
Consider the 2-dimensional complex distribution 
in $TM\otimes\mathbb{C}$ given by
$D=\{o^{A}\beta^{A'} \;|\; \beta^{A'}\in\mathbb{S}'\}$. 
The condition for this to be involutive (i.e. $[D,D]\subset D$) is 
the shear-free equation $o^{A}o^{B}\nabla_{AA'}o_{B}=0$ 
(cf. \cite[Section 7.3]{PR2}),
which is certainly satisfied if \eqref{PS} holds. 
This implies that there exist complex 2-surfaces in the 
complexified space-time $\mathbb{C}M$, called 
$\beta$-surfaces, such that their tangent bundle is $D$.
Analogously, the distribution $\bar{D}=\{\bar{o}^{A'}\alpha^{A} \;|\; 
\alpha^{A}\in\mathbb{S}\}$ is involutive, and is the tangent 
bundle to a different kind of complex 2-surfaces in 
$\mathbb{C}M$, called $\alpha$-surfaces. 
Let us focus on the former.
The $\beta$-surfaces are labelled by two complex coordinates 
$(v,\zeta)$ that are constant on them, namely 
$o^{A}\nabla_{AA'}v=0$, $o^{A}\nabla_{AA'}\zeta=0$ 
(see \cite[Lemma (7.3.15)]{PR2}).
From these two equations we deduce that there are 
two spinor fields, say $v_{A'},\bar{\iota}_{A'}$, such that 
$\nabla_{AA'}v=o_{A}v_{A'}$, 
$\nabla_{AA'}\zeta=o_{A}\bar{\iota}_{A'}$. 
Since $\ell_{a}=o_{A}\bar{o}_{A'}$ is covariantly constant, it is in particular closed, so we can take $v_{A'}=\bar{o}_{A'}$. 
So $(v,\zeta)$ are defined by
\begin{equation}
 {\rm d}v = o_{A}\bar{o}_{A'}{\rm d}x^{AA'}, \qquad 
 {\rm d}\zeta = o_{A}\bar{\iota}_{A'}{\rm d}x^{AA'}.
  \label{CoordConstBeta}
\end{equation}
We see that $v$ is real, whereas $\zeta$ is complex. 
They are functionally independent, which means that
$\bar{o}_{A'}\bar{\iota}^{A'}=N$ for some scalar field $N\neq0$.

\smallskip
From the condition ${\rm d}^{2}\zeta=0$ we deduce that 
$o^{A}\nabla_{AA'}\bar{\iota}_{B'}=0$. 
Therefore, $o^{A}\nabla_{AA'}N=0$, 
which implies that $N$ is a holomorphic function of $v,\zeta$ 
(i.e. ${\rm d}N$ is a linear combination of ${\rm d}v$ and 
${\rm d}\zeta$), so it can be set to 1 by a coordinate 
transformation $\zeta \to \zeta'(v,\zeta)$. 
We drop the prime and denote again by $\zeta$ 
the new coordinate, with $\bar{o}_{A'}\bar\iota^{A'}=1$.

\smallskip
Notice that $v,\bar{\zeta}$ satisfy $\bar{o}^{A'}\nabla_{AA'}v=0$, 
$\bar{o}^{A'}\nabla_{AA'}\bar{\zeta}=0$, so these scalars are 
constant on $\alpha$-surfaces. This is a generic feature of 
Lorentz signature: spinor complex conjugation interchanges 
$\alpha$- and $\beta$-surfaces.

\begin{figure}
\centering
\tdplotsetmaincoords{70}{110}

\begin{tikzpicture}[tdplot_main_coords]

\draw[fill=blue,opacity=0.2] (-3,1,-2) -- (-3,-1,2) -- (3,-1,2) -- (3,1,-2) -- cycle;
\draw[fill=red,opacity=0.2] (-3,-1,-2) -- (-3,1,2) -- (3,1,2) -- (3,-1,-2) -- cycle;
\draw[thick](-3,0,0)--(3,0,0);

\node[anchor=south west,align=center] (line) at (4,4,2) 
{real light ray $\gamma\subset M$};

\node[anchor=south west,align=center] at (3,3.1,3.3) {\textcolor{red}{$\beta$-surface $\subset\mathbb{C}M$} \\ \textcolor{red}{$(v,\zeta)=$constant}};
\node[anchor=south west,align=center] at (0,-4,1.5) {\textcolor{blue}{$\alpha$-surface $\subset\mathbb{C}M$} \\ \textcolor{blue}{$(v,\bar{\zeta})=$constant}};

\draw[-latex] (line) to[out=180,in=75] (-2,0,0.05);

\draw[->,color=violet] (3,0,0) -- (0,0,0) node[right] {$\partial_{u}$};
\draw[->,color=violet] (3,0,0) -- (3,-0.5,1) node[left] {$\partial_{\zeta}$};
\draw[->,color=violet] (3,0,0) -- (3,0.5,1) node[left] {$\partial_{\bar{\zeta}}$};

\end{tikzpicture}
\caption{An $\alpha$-surface, a $\beta$-surface, and a real 
(Lorentzian) space-time intersect in a real light ray $\gamma$, that has tangent vector $\ell^{a}=o^{A}\bar{o}^{A'}=\partial^{a}_{u}$. 
The coordinate system defined by these twistor surfaces coincides, 
in the pp-wave case, with Brinkmann coordinates.}
\label{Fig:betasurface}
\end{figure}
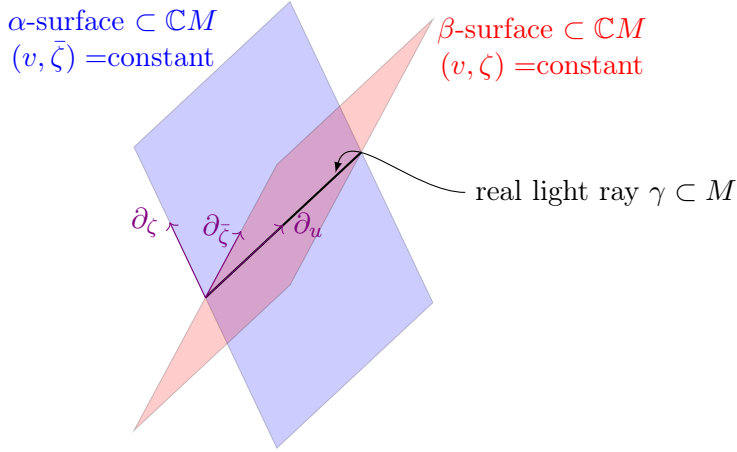

Using some of the previous identities, one can show that
the vector fields $o^{A}\bar{o}^{A'}$ and $o^{A}\bar\iota^{A'}$ 
commute, so the two scalar fields $u,w$ defined by
$\partial_{u} = o^{A}\bar{o}^{A'}\partial_{AA'}$, 
$\partial_{w} = o^{A}\bar\iota^{A'}\partial_{AA'}$ 
are functionally independent. 
We see that $u$ is real and $w$ is complex. 
These are coordinates along the $\beta$-surfaces.
The coordinate $w$ is however not functionally independent of 
$(v,\bar{\zeta})$, since a short calculation gives 
$\partial_{\bar{\zeta}}=-o^{A}\bar{\iota}^{A'}\partial_{AA'}$.
Summarizing, we have
\begin{equation}
 \partial_{u} = o^{A}\bar{o}^{A'}\partial_{AA'}, \qquad
 \partial_{\bar{\zeta}} = -o^{A}\bar\iota^{A'}\partial_{AA'}.
 \label{coordinateVFs}
\end{equation}
So $\alpha$- and $\beta$-surfaces give a coordinate system 
$(u,v,\zeta,\bar{\zeta})$ for $M$, that we illustrate in 
Fig. \ref{Fig:betasurface}.
For pp-waves these are simply Brinkmann coordinates, 
so the interpretation is known: 
the integral curves of $\ell^{a}=o^{A}\bar{o}^{A'}$ are the 
rays of the wave and $u$ is an affine parameter along them, 
the hypersurfaces $v={\rm constant}$ are `wave surfaces',
and the real and imaginary parts of $\zeta$ are coordinates 
transverse to the direction of propagation of the wave.

\medskip
With the above information, the structure of the metric can 
be deduced from the expression 
$g= \epsilon_{AB}\bar{\epsilon}_{A'B'}{\rm d}x^{AA'}\otimes{\rm d}x^{BB'}$, by replacing
$\epsilon_{AB}=o_{A}\iota_{B}-\iota_{A}o_{B}$, its complex 
conjugate, and the definition of the coordinates \eqref{CoordConstBeta}.
When doing this computation, one finds that the only 
piece of information missing at this point is an expression for 
the 1-form $\iota_{A}\bar{\iota}_{A'}{\rm d}x^{AA'}$.
This can be obtained as follows. 
For any function $f$, we have 
${\rm d}f=(\nabla_{AA'}f){\rm d}x^{AA'}$. 
Using the identities 
$\nabla_{AA'}=\delta_{A}^{B}\bar{\delta}_{A'}^{B'}\nabla_{BB'}$ 
and $\delta^{B}_{A}=o_{A}\iota^{B}-\iota_{A}o^{B}$, 
together with definitions \eqref{CoordConstBeta}, 
\eqref{coordinateVFs}, and putting $f=u$, we get 
\begin{align*}
 \iota_{A}\bar\iota_{A'}{\rm d}x^{AA'} 
 = {\rm d}u - (\iota^{A}\bar{\iota}^{A'}\nabla_{AA'}u){\rm d}v.
\end{align*}
The expression \eqref{ppwavemetric} for the metric then 
follows straightforwardly, by defining the real scalar field
\begin{equation}
 H:=-2\iota^{A}\bar\iota^{A'}\nabla_{AA'}u. \label{H}
\end{equation}
This function $H$ represents the wave profile, 
and the case $H=0$ reduces to Minkowski space-time.

We also notice that in the coordinate system
 $(u,v,\zeta,\bar{\zeta})$,
the wave operator acting on an arbitrary scalar field $\varphi$ is
\begin{equation}
 \Box\varphi = 2(\varphi_{uv}-\varphi_{\zeta\bar{\zeta}}) 
 - H\varphi_{uu}.
 \label{Boxppwave}
\end{equation}

\medskip
There is a natural spin frame:
the parallel spinor $o_{A}$, and the spinor $\iota_{A}$ used 
in \eqref{CoordConstBeta}. 
The associated connection 1-form has only one non-trivial 
component:
\begin{equation}
 \nabla_{AA'}\iota_{B} = -\kappa' o_{A}o_{B}\bar{o}_{A'}, 
 \label{nablaiota}
\end{equation}
where 
$\kappa':=-\iota^{A}\iota^{B}\bar\iota^{B'}\nabla_{BB'}\iota_{A}$.
In order to show this, notice first that, from \eqref{CoordConstBeta} and ${\rm d}^{2}\bar{\zeta}=0$, 
it follows that $\bar{o}^{A'}\nabla_{AA'}\iota_{B}=0$. 
In addition, from \eqref{CoordConstBeta} 
one deduces that $o_{A}=-\bar{o}^{A'}\nabla_{AA'}\zeta$ 
and $\iota_{A}=\bar\iota^{A'}\nabla_{AA'}\bar{\zeta}$.
These identities can then be used to show that 
$o^{A}\nabla_{AA'}\iota_{B}=0$, so \eqref{nablaiota} follows.
In terms of $H$, the expression for $\kappa'$ is
\begin{equation}
 \kappa' = \tfrac{1}{2}H_{\zeta}. \label{kappa'}
\end{equation}
This can be shown by using \eqref{H}, 
which gives $ o^{A}\nabla_{AA'}H
 = 2\iota^{B}\bar\iota^{B'}\nabla_{BB'}\bar{\iota}_{A'}$.

For the curvature, eq. \eqref{PS} implies $[\nabla_{a},\nabla_{b}]o^{D}=0$, so it follows that $\Lambda=0$,
$\Phi_{ABA'B'}=\Phi_{22}o_{A}o_{B}\bar{o}_{A'}\bar{o}_{B'}$, 
and $\Psi_{ABCD} = \Psi_{4}o_{A}o_{B}o_{C}o_{D}$,
where $\Phi_{22}$ and $\Psi_{4}$ are defined by contractions with $\iota^{A},\bar{\iota}^{A'}$ on the left-hand sides.
Expressions for $\Phi_{22}$ and $\Psi_{4}$ in terms of 
$H$ can be deduced, for example, by using 
the Newman-Penrose equations and the fact that all spin coefficients except $\kappa'$ vanish: 
equations $(4.11.12)(a')$ and $(4.11.12)(b')$ in \cite{PR1} 
give $\Phi_{22}=-\delta\kappa'$ and $\Psi_{4}=-\delta'\kappa'$ 
(in Newman-Penrose notation).
Using then $\delta=-\partial_{\bar{\zeta}}$,
$\delta'=-\partial_{\zeta}$ 
(which follow from \eqref{coordinateVFs}), and \eqref{kappa'}, 
one obtains the expressions 
\eqref{Riccippwave}, \eqref{Weylppwave}.

\subsection{Complex space-times}
\label{Sec:Complexppwaves}

We now consider a complex space-time with a parallel spinor.
Here, we are referring to a genuinely 
complex 4-manifold, not to a complexified pp-wave 
space-time. See \cite[Section 6.9]{PR2} for 
the distinction between `complex' and `complexified' space-time.
In the complexified case we still have two parallel spinors 
of opposite chirality ($o^{A}$ and $\bar{o}^{A'}$ become 
independent but they are both parallel), 
whereas in the genuinely complex case 
we only retain one parallel spinor.

\smallskip
The situation is very similar to the case of split signature.
This is because in that case, 
spinors are real and the two chiralities are independent.
The following result is for complex space-times, but
if one replaces ``complex'' by ``real'' everywhere then exactly 
the same holds true for split signature metrics, as was 
already shown in \cite{Dunajski, DP}: 

\begin{proposition}[See \cite{Dunajski, DP}]
\label{Prop:Complexppwave}
Let $(\mathbb{C}M,g^{\mathbb{C}}_{ab})$ be a complex 
space-time with a non-trivial parallel spinor field $o^{A}$.
Then there exist, locally, a complex coordinate system
$(u,v,\zeta,w)$ and a complex scalar field $\Theta$ such that 
the metric is
\begin{equation}
g^{\mathbb{C}} = 2({\rm d}u{\rm d}v+{\rm d}\zeta{\rm d}w) 
 -2\Theta_{ww}{\rm d}v^{2} +4\Theta_{wu}{\rm d}v{\rm d}\zeta 
 -2 \Theta_{uu}{\rm d}\zeta^{2}.
 \label{Complexppwavemetric}
\end{equation}
The Ricci scalar vanishes, and the rest of the curvature is given by
\begin{align}
 \Phi_{ABA'B'} ={}& o_{A}o_{B}\tn_{A'}\tn_{B'}f, 
 \label{RicciComplexppwave} \\
 \Psi_{ABCD} ={}& -\tfrac{1}{2}o_{A}o_{B}o_{C}o_{D}\Box f, 
 \label{ASDWeylComplexppwave} \\
 \tilde{\Psi}_{A'B'C'D'} ={}&-\tn_{A'}\tn_{B'}\tn_{C'}\tn_{D'}\Theta,
 \label{SDWeylComplexppwave}
\end{align}
where $\tn_{A'}:=o^{A}\nabla_{AA'}$, $\Box$ is the wave operator 
associated to \eqref{Complexppwavemetric}, and 
\begin{equation}
 f = \Theta_{uv}+\Theta_{\zeta w} + \Theta_{uu}\Theta_{ww} 
 -\Theta_{uw}^{2}.
\end{equation}
\end{proposition}

In coordinates, the wave operator $\Box$ 
acting on an arbitrary scalar field $\varphi$ is
\begin{equation}
 \Box\varphi=2(\varphi_{uv}+\varphi_{w\zeta}
 +\Theta_{uu}\varphi_{ww}+\Theta_{ww}\varphi_{uu}
 -2\Theta_{uw}\varphi_{uw}). 
 \label{BoxComplexppwave}
\end{equation}

From Prop. \ref{Prop:Complexppwave}
we see that the vacuum Einstein equations are now 
more complicated than in the pp-wave case: 
the Ricci-flat condition is equivalent to 
$\tn_{A'}\tn_{B'}f=0$, which in the coordinate system of 
the proposition reads $f_{uu}=f_{uw}=f_{ww}=0$. 
The solution to this is 
$f=p(v,\zeta)u+q(v,\zeta)w+r(v,\zeta)$, 
where $p,q,r$ are arbitrary functions of $(v,\zeta)$,
so in terms of the ``potential'' $\Theta$, the Einstein equations 
are 
\begin{equation}
 \Theta_{uv}+\Theta_{\zeta w} + \Theta_{uu}\Theta_{ww} 
 -\Theta_{uw}^{2} 
 = p(v,\zeta)u+q(v,\zeta)w+r(v,\zeta),
 \label{HH}
\end{equation}
see \cite{Dunajski}. This is a very special case of 
the hyper-heavenly equation of Pleba\'nski and Robinson 
\cite{PlebanskiRobinson}. 
The non-trivial right hand side in \eqref{HH} (i.e. $f\neq 0$)
complicates the 
analysis of the integrability properties of this equation.
The special case $f\equiv 0$ is Pleba\'nski's second 
heavenly equation, and notice from \eqref{ASDWeylComplexppwave} that this case gives 
a self-dual (half-flat) space, which is an integrable system 
by virtue of the non-linear graviton twistor construction 
of Penrose.

\medskip
It is useful to briefly discuss the structures involved 
in the derivation of the result of Prop. \ref{Prop:Complexppwave}. 
As in section \ref{Sec:ppwaves},
the condition $\nabla_{AA'}o^{B}=0$ implies that the distribution 
$D=\{o^{A}\beta^{A'} \;|\; \beta^{A'}\in\mathbb{S}'\}$ is involutive, 
and this gives origin to $\beta$-surfaces in $\mathbb{C}M$, 
which are labelled 
by two complex coordinates $v,\zeta$ defined 
by $o^{A}\nabla_{AA'}v=0=o^{A}\nabla_{AA'}\zeta$.
Unlike the Lorentzian case, there are no $\alpha$-surfaces now.
In addition, both coordinates $v,\zeta$ are now complex. 
There are two independent spinor fields 
$v_{A'},\zeta_{A'}$, with $v_{A'}\zeta^{A'}=N\neq0$, such that 
\begin{align}
 {\rm d}v=o_{A}v_{A'}{\rm d}x^{AA'}, \qquad
 {\rm d}\zeta = o_{A}\zeta_{A'}{\rm d}x^{AA'}.
 \label{complexcoordinates}
\end{align}
From the conditions ${\rm d}^{2}v=0={\rm d}^{2}\zeta$, 
it follows that $\tn_{A'}v_{B'}=\tn_{A'}\zeta_{B'} = 0$, which 
give $\tn_{A'}N=0$, so we can set $N\equiv 1$ 
by a coordinate transformation. 

Using the above information, a short calculation shows that 
the vector fields $o^{A}v^{A'}$ and $o^{A}\zeta^{A'}$ commute, 
so the two complex scalar fields $u$, $w$ defined by 
\begin{equation}
 o^{A}v^{A'}\partial_{AA'} = \partial_{u}, \qquad
 o^{A}\zeta^{A'}\partial_{AA'} = \partial_{w}
 \label{complexcoordinates2}
\end{equation}
are functionally independent, and correspond to complex 
coordinates along the $\beta$-surfaces.
Thus, we see again that twistor surfaces produce a 
natural coordinate system $(v,\zeta,u,w)$ for $\mathbb{C}M$: 
these are the coordinates used in Prop. \ref{Prop:Complexppwave}, 
and they generalize the Brinkmann coordinates 
of the pp-wave case of Prop. \ref{Prop:ppwave}.

The structure \eqref{Complexppwavemetric} of the metric can 
be deduced in a similar way to what we did in section \ref{Sec:ppwaves}:
there is a flat (complex) metric $\eta_{ab}$ and 
a symmetric spinor field $H_{A'B'}$ such that
\begin{equation}
 g_{ab} = \eta_{ab} + o_{A}o_{B}H_{A'B'}.
 \label{AbstractComplexMetric}
\end{equation}
The components of $H_{A'B'}$ generalize the pp-wave 
profile function $H$ \eqref{H}.
In addition, a short calculation shows that $H_{A'B'}$ satisfies 
$\tn^{A'}H_{A'B'}=0$, so there exists a scalar field $\Theta$ 
(see Remark \ref{Remark:Potentials} below) such that
\begin{equation}
 H_{A'B'} = -2\tn_{A'}\tn_{B'}\Theta. 
 \label{HAB}
\end{equation} 
The equations \eqref{AbstractComplexMetric}-\eqref{HAB} 
give a coordinate-free expression for \eqref{Complexppwavemetric}. 
The Einstein equation \eqref{HH} is, in coordinate-free terms:
\begin{equation}
 \Box\Theta 
 -2 (\tn_{A'}\tn_{B'}\Theta)(\tn^{A'}\tn^{B'}\Theta) = f, 
 \qquad \tn_{A'}\tn_{B'}f=0.
 \label{AbstractHH}
\end{equation}

\section{Perturbations}
\label{Sec:Perturbations}

We will now study real gravitational perturbations of 
a real, Lorentzian, vacuum pp-wave space-time, and 
connections with complex space-times.
We recall that the structure of the background pp-wave 
space-time is described in Prop. \ref{Prop:ppwave}: 
one has Brinkmann coordinates $(u,v,\zeta,\bar\zeta)$ 
defined by $\alpha$- and $\beta$-surfaces, 
the spinor field $o^{A}$ is parallel, the spinor $\iota^{A}$
is defined in eq. \eqref{CoordConstBeta}, and 
all the information of the geometry is encoded in 
the real scalar field $H$.
In addition, the vacuum condition for the background 
implies that $H_{\zeta\bar\zeta}=0$.

\subsection{Main result}
\label{Sec:MainResult}

\begin{theorem}\label{Thm:Main}
Let $(M,g_{ab})$ be a vacuum pp-wave space-time, 
eq. \eqref{ppwavemetric} with $H_{\zeta\bar\zeta}=0$.
For any real metric perturbation $h_{ab}$ 
satisfying the linearized Einstein vacuum equations, 
there exist, locally, 
a real vector field $V_{a}$ and a complex scalar field 
$\Phi$, such that $h_{ab}$ can be written as
\begin{equation}
 h_{ab} = 2{\rm Re}(h^{\rm H}_{ab}) + \nabla_{a}V_{b} + 
 \nabla_{b}V_{a}
\end{equation}
where $h^{\rm H}_{ab}$ is given by
\begin{equation}
 h^{\rm H}_{ab} = o_{A}o_{B}\tn_{A'}\tn_{B'}\Phi 
 =\Phi_{\bar{\zeta}\bar{\zeta}}\ell_{a}\ell_{b} 
 +2\Phi_{\bar{\zeta}u}\ell_{(a}m_{b)} + \Phi_{uu}m_{a}m_{b}
  \label{Hertz}
\end{equation}
with $\tn_{A'}=o^{A}\nabla_{AA'}$,
and $\Phi$ satisfies the wave equation
\begin{equation}
 \Box\Phi = 0, \label{WaveEquation}
\end{equation}
where $\Box$ is the wave operator associated to $g_{ab}$, 
eq. \eqref{Boxppwave}.
\end{theorem}

We prove this result in section \ref{Sec:Proof} below.
Note that in tensor terms, 
the tensor field \eqref{Hertz} can also be written as 
\begin{equation}
h^{\rm H}_{ab} = 
\nabla_{c}\nabla_{d}[\mathcal{H}_{(a}{}^{cd}{}_{b)}\Phi], 
\qquad 
\mathcal{H}_{abcd} = 4\ell_{[a}m_{b]}\ell_{[c}m_{d]}.
\end{equation}
This is the usual expression for a Hertz/Debye potential 
in perturbation theory, particularized to the special background 
of a pp-wave. (The superscript ``${\rm H}$'' is from ``Hertz''.)

\smallskip
Combining the result of theorem \ref{Thm:Main} 
with the discussion of section \ref{Sec:Complexppwaves}, 
we see some sort of correspondence between a {\em real linear} 
problem and a {\em complex non-linear} one: 
modulo gauge, the linearized gravity problem for real 
pp-wave space-times would seem to be a ``linear 
version'' (see below) of the structure of a complex space with 
a parallel spinor. Furthermore, as observed in 
section \ref{Sec:Complexppwaves}, the non-linear structures 
can actually be understood in a {\em real} context, by going 
to a real space with a split signature metric.

\smallskip
Note that this correspondence can be established only 
{\em after} one proves theorem \ref{Thm:Main}:
we want to show that {\em there exists} a scalar potential for 
the gravitational perturbation, while in a linear version of 
\eqref{AbstractComplexMetric}--\eqref{AbstractHH}
one is already assuming that a potential exists. 
Actually, a closer look at the linear version of
\eqref{AbstractComplexMetric}--\eqref{AbstractHH}
reveals that the situation is subtle:

\begin{itemize}
\item By ``linear version'' we mean that, in the complex metric
\eqref{AbstractComplexMetric}-\eqref{HAB}
and in the complex Einstein equations \eqref{HH},
one formally 
replaces $\Theta$ by $\Theta + \varepsilon\Phi$  
(where $\varepsilon$ is a parameter),
and one keeps only linear terms in $\varepsilon$. 
Then the perturbation to the background complex metric 
is exactly \eqref{Hertz}, and the scalar field $\Phi$ 
would seem to satisfy the wave equation $\Box\Phi=0$. 
In addition, using the general expression \cite[Eq. (5.7.15)]{PR1}
for the perturbed Weyl spinor, it is not difficult to show 
that for the complex perturbation \eqref{Hertz} one has
\begin{equation}
  \dot{\Psi}_{ABCD}[h^{\rm H}] = 
  \tfrac{1}{8}o_{A}o_{B}o_{C}o_{D}\Box\Box\Phi,
  \qquad
 \dot{\tilde{\Psi}}_{A'B'C'D'}[h^{\rm H}]=
 \tfrac{1}{2}\tn_{A'}\tn_{B'}\tn_{C'}\tn_{D'}\Phi, 
\end{equation}
which resemble \eqref{ASDWeylComplexppwave}, 
\eqref{SDWeylComplexppwave}.
\item However, the linear version of the complex 
Einstein equations is the fourth order equation 
$\tn_{A'}\tn_{B'}\Box\Phi=0$, see \eqref{RicciComplexppwave}. 
Analogously to the discussion around eq. \eqref{HH},
this implies that $\Box\Phi=F$, where $F$ is 
a function such that $F_{uu}=F_{uw}=F_{ww}=0$, 
so one does not really get the homogeneous wave equation 
for $\Phi$. 
We will encounter a similar issue in our proof of 
theorem \ref{Thm:Main} below, where we will show that 
one can get rid of inhomogeneous terms by considering 
gauge transformations.
\item Even if one manages to get the homogeneous 
wave equation, the background real and complex geometries, 
eqs. \eqref{ppwavemetric} and \eqref{Complexppwavemetric} 
respectively, are different, which means that the wave 
equations, while {\em formally} equal, are different in practice.
Explicitly, the wave operators of the real and complex 
geometries are given by equations \eqref{Boxppwave} 
and \eqref{BoxComplexppwave}.
\end{itemize}

\subsection{Proof of theorem \ref{Thm:Main}}
\label{Sec:Proof}

\subsubsection{Preliminaries}

We consider a smooth mono-parametric family of real space-times 
$(M,g_{ab}(\varepsilon))$, where $g_{ab}\equiv g_{ab}(0)$ is the 
background space-time and is assumed to satisfy
the vacuum Einstein equations $R_{ab}=0$. 
The background Levi-Civita connection is denoted by $\nabla_{a}$, 
and the linearization of the metric is 
$h_{ab}=\frac{\rm d}{{\rm d}\varepsilon}|_{\varepsilon=0} 
[g_{ab}(\varepsilon)]$.
The linearizations of the Ricci tensor and of the curvature 
scalar are linear operators acting on $h_{ab}$. 
They will be denoted by $\dot{R}_{ab}[h]$ and $\dot{R}[h]$
respectively, and explicit expressions for them are 
(see e.g. \cite{Wald})
\begin{align}
 \dot{R}_{ab}[h]={}&-\tfrac{1}{2}\Box h_{ab} 
 -\tfrac{1}{2}\nabla_{a}\nabla_{b}(g^{cd}h_{cd}) 
 +\tfrac{1}{2}\nabla^{c}\nabla_{a}h_{bc}
 +\tfrac{1}{2}\nabla^{c}\nabla_{b}h_{ac},
 \label{linRicciGeneric} \\
 \dot{R}[h] ={}& \nabla^{a}\nabla^{b}h_{ab} 
 - \Box(g^{ab}h_{ab}). \label{linRicciScalarGeneric}
\end{align}

Calculations are greatly simplified by using spinors. 
We emphasize that we do not perturb spinors themselves, 
we just use the spinor structure of the background space-time. 
Since all perturbations are tensor fields, they can be 
written in spinor language, using the usual dictionary 
between tensor indices and pairs of spinor indices (see 
section \ref{Sec:Preliminaries}).
For example, using the background Levi-Civita 
connection $\nabla_{a}=\nabla_{AA'}$, for the perturbed 
Ricci tensor we can write 
$\dot{R}_{ab}[h]\equiv\dot{R}_{ABA'B'}[h]$, with 
\begin{equation}
\dot{R}_{ABA'B'}[h] = -\frac{1}{2} \left[ \Box h_{ABA'B'} + 
\nabla_{AA'}\nabla_{BB'}(g^{cd}h_{cd}) - 
\nabla^{CC'}\nabla_{AA'}h_{BB'CC'} - 
\nabla^{CC'}\nabla_{BB'}h_{AA'CC'} \right]. 
\label{LinRicciSpinor}
\end{equation}
Notice that this does not mean that we are linearizing a 
spinor field. 
In this work, the meaning of ``perturbed field'' is the 
ordinary one in perturbation theory in general relativity 
(see e.g. \cite{Wald}).

\medskip
Let $\{o^{A},\iota^{A}\}$ be a spin frame for the background 
space-time, $\epsilon_{AB}o^{A}\iota^{B}=1$. 
For the calculations in this section, it is useful to 
define the operators
\begin{equation}
 \tn_{A'}:=o^{A}\nabla_{AA'}, \quad \nabla_{A'}:=\iota^{A}\nabla_{AA'}, \quad \bn_{A}:=\bar{o}^{A'}\nabla_{AA'}, \quad \nabla_{A}:=\bar{\iota}^{A'}\nabla_{AA'}. 
 \label{ProjectedDerivatives}
\end{equation}
For the particular case of a pp-wave background, 
from the discussion of section \ref{Sec:ppwaves} we have
$\tn_{A'}o^{B}=\nabla_{A'}o^{B}=0$, $\tn_{A'}\iota^{B}=0$ 
and $\nabla_{A'}\iota^{B} = -\kappa'\bar{o}_{A'}o^{B}$.

\subsubsection{The radiation gauge}
\label{Sec:RadiationGauge}

As is well-known, diffeomorphism invariance in 
general relativity implies that in linearized gravity, 
any metric perturbation $h_{ab}$ is physically 
equivalent to $h_{ab}+K[\xi]_{ab}$, where 
$K[\xi]_{ab} = \nabla_{a}\xi_{b}+\nabla_{b}\xi_{a}$ 
and $\xi_{a}$ is arbitrary.
For a vacuum background, it identically holds
$\dot{R}_{ab}[K[\xi]] \equiv 0$ for any $\xi_{a}$.
For a pp-wave, in appendix \ref{App:gauge} 
we give explicit expressions for the components of $K[\xi]_{ab}$.

\smallskip
For a background space-time possessing a null vector $\ell^{a}$  
associated to a geodesic shear-free congruence 
(which is certainly the case for the pp-waves studied 
in this work), one can impose (see \cite{Whiting}) the 
so-called {\em radiation gauge}:
\begin{equation}
 \ell^{a}h_{ab} = 0, \qquad g^{ab}h_{ab} = 0. 
 \label{RadiationGauge}
\end{equation}
A short calculation then shows that in terms of a null tetrad 
$\{\ell_{a},n_{a},m_{a},\bar{m}_{a}\}$, it holds
\begin{equation}
h_{ab} = h_{nn}\ell_{a}\ell_{b} - 2h_{n\bar{m}}\ell_{(a}m_{b)} - 2h_{nm}\ell_{(a}\bar{m}_{b)}+h_{\bar{m}\bar{m}}m_{a}m_{b} 
+h_{mm}\bar{m}_{a}\bar{m}_{b}
\label{perturbationRG}
\end{equation}
where $h_{nn}=n^{a}n^{b}h_{ab}$, etc.
Replacing the expression \eqref{nulltetrad} for the null vectors, 
in spinor language we get
\begin{equation*}
 h_{ab} = o_{A}o_{B}\mathring{X}_{A'B'} + 
 \bar{o}_{A'}\bar{o}_{B'}\bar{\mathring{X}}_{AB},
\end{equation*}
where 
\begin{equation*}
 \mathring{X}_{A'B'} = 
 \tfrac{1}{2}h_{nn}\bar{o}_{A'}\bar{o}_{B'} 
 - 2h_{n\bar{m}}\bar{o}_{(A'}\bar\iota_{B')} 
 + h_{\bar{m}\bar{m}}\bar{\iota}_{A'}\bar{\iota}_{B'}.
\end{equation*}
Now, let $\psi$ be an arbitrary real scalar field. Then 
we have, trivially,
\begin{align}
\nonumber h_{ab} ={}& o_{A}o_{B}\mathring{X}_{A'B'} + 
 \bar{o}_{A'}\bar{o}_{B'}\bar{\mathring{X}}_{AB}
 +i\psi o_{A}o_{B}\bar{o}_{A'}\bar{o}_{B'} 
 -i\psi o_{A}o_{B}\bar{o}_{A'}\bar{o}_{B'} \\
={}& o_{A}o_{B} \underbrace{ (\mathring{X}_{A'B'}
 +i\psi \bar{o}_{A'}\bar{o}_{B'})}_{=: X_{A'B'}}
 +\bar{o}_{A'}\bar{o}_{B'} 
 \underbrace{ (\bar{\mathring{X}}_{AB}
 -i\psi o_{A}o_{B})}_{=\bar{X}_{AB}},
 \label{ooX0}
\end{align}
so the tensor field \eqref{perturbationRG} is 
\begin{equation}
 h_{ab} = \gamma_{ab} + \bar{\gamma}_{ab}, 
 \qquad 
 \gamma_{ab} = o_{A}o_{B}X_{A'B'}. \label{ooX} 
\end{equation}
The reason for including the arbitrary scalar field $\psi$ 
will become clear in the next section.
The relation between the components of 
$X_{A'B'}$ and the components in \eqref{perturbationRG} is
\begin{equation}
 X_{0'0'}=h_{\bar{m}\bar{m}}, \qquad 
 X_{0'1'}= h_{n\bar{m}}, \qquad 
 X_{1'1'} =\tfrac{1}{2}h_{nn} + i\psi.
\end{equation}

It is important to note that the conditions \eqref{RadiationGauge} do not exhaust the gauge freedom. 
In Appendix \ref{App:ResidualGauge} we analyse 
the residual gauge transformations under which 
\eqref{RadiationGauge} is preserved. 
This plays an important role below.

\subsubsection{Potentials}
\label{Sec:Potentials}

We now assume that we are given a real metric perturbation
$h_{ab}$ in radiation gauge, eq. \eqref{ooX}, 
that satisfies the linearized Einstein vacuum equations:
\begin{equation}
 \dot{R}_{ABA'B'}[h] =
 \dot{R}_{ABA'B'}[\gamma] +\dot{R}_{ABA'B'}[\bar{\gamma}]
 =0.
 \label{LinearizedEinsteinEq}
\end{equation}
Notice that this equation does not imply that 
$\dot{R}_{ABA'B'}[\gamma]$ vanishes.
Using \eqref{LinRicciSpinor}, \eqref{ProjectedDerivatives}, 
and \eqref{ooX}, after some calculations we find 
the following expressions for the linearized Ricci tensor and Ricci
scalar of the tensor field $\gamma_{ab} = o_{A}o_{B}X_{A'B'}$:
\begin{subequations}\label{RicciooX}
\begin{align}
 \dot{R}[\gamma] ={}& \tn^{A'}\tn^{B'}X_{A'B'}, 
 \label{RSooX} \\
 \dot{R}_{ABA'B'}[\gamma] ={}& -\frac{1}{2} \left[ 
 o_{A}o_{B}\Box X_{A'B'} + o_{B}\tn^{A'}\nabla_{AA'}X_{B'C'}
 +o_{A}\tn^{A'}\nabla_{BB'}X_{A'C'}
 \right]. \label{RTooX}
\end{align}
\end{subequations}
For the complex conjugate $\bar{\gamma}_{ab}=
\bar{o}_{A'}\bar{o}_{B'}\bar{X}_{AB}$, 
the corresponding formulas are obtained by simply 
taking the complex conjugate of the above.
Note that, regardless of \eqref{LinearizedEinsteinEq}, 
it follows immediately that 
\begin{equation}
 o^{A}o^{B}\dot{R}_{ABA'B'}[\gamma] \equiv 0, \qquad 
 \bar{o}^{A'}\bar{o}^{B'}\dot{R}_{ABA'B'}[\bar{\gamma}] \equiv 0.
 \label{ooRicci}
\end{equation}

From \eqref{RSooX} and its complex conjugate,
we find
\begin{align}
\nonumber \dot{R}[h] ={}& \tn^{A'}\tn^{B'}X_{A'B'} 
 + \bn^{A}\bn^{B}\bar{X}_{AB} \\
 ={}&  (\tn^{A'}\tn^{B'}\mathring{X}_{A'B'}  + i\partial^{2}_{u}\psi)
 +(\bn^{A}\bn^{B}\overline{\mathring{X}}_{AB}- i\partial^{2}_{u}\psi)
 =0,
 \label{RicciScalar=0}
\end{align}
where in the second line we used the definition of 
$X_{A'B'}$ given in eq. \eqref{ooX0}.
We now see the reason for including the arbitrary scalar 
field $\psi$: since it is free, we can choose it so as to satisfy
\begin{equation}
 \partial^{2}_{u}\psi = i \tn^{A'}\tn^{B'}\mathring{X}_{A'B'},
\end{equation}
or more explicitly:
\begin{equation}
  \partial^{2}_{u}\psi = 
  i \left[ \partial^{2}_{\bar{\zeta}}h_{\bar{m}\bar{m}} 
 +2\partial_{u}\partial_{\bar{\zeta}} h_{n\bar{m}}
 +\tfrac{1}{2}\partial^{2}_{u}h_{nn} \right].
 \label{EqForpsi}
\end{equation}
(One still has the freedom $\psi \to \psi + \chi$ with 
$\partial^{2}_{u}\chi=0$, but we will not need this.)
The choice \eqref{EqForpsi} of $\psi$ has the consequence that 
\begin{equation}
\tn^{A'}\tn^{B'}X_{A'B'} = 0, \label{d2X}
\end{equation}
which implies that there exists, locally, 
a spinor field $Y_{A'}$ such that 
\begin{equation}
 X_{A'B'} = \tn_{(A'}Y_{B')}. \label{Y}
\end{equation}

\begin{remark}\label{Remark:Potentials}
The argument for deducing \eqref{Y} from \eqref{d2X} 
is essentially a variant of the argument given by Penrose in 
\cite[Section 4]{Penrose65}. 
It is always true locally, and it can be extended globally to a 
region that has vanishing first and second homotopy groups. 
As explained by Penrose, this topological restriction accounts for 
the impossibility of finding {\em global} potentials in certain cases, 
such as for Coulomb fields.
\end{remark}

In coordinate terms, the above means that, by virtue of 
the vanishing of the perturbed Ricci scalar of $h_{ab}$, 
and by choosing $\psi$ in the form \eqref{EqForpsi}, 
one can locally find two fields 
$Y_{0'}$, $Y_{1'}$ (which are the components of a spinor 
field $Y_{A'}$ in the spin frame $\{\bar{o}_{A'},\bar{\iota}_{A'}\}$)
such that 
\begin{subequations}\label{RelationsYmetric}
\begin{align}
 \partial_{u}Y_{0'} ={}& 
 h_{\bar{m}\bar{m}}, \label{RelationYmetric1} \\
 \partial_{u}Y_{1'} - \partial_{\bar{\zeta}}Y_{0'} 
 ={}& 2h_{n\bar{m}}, \label{RelationYmetric2} \\
 \partial_{\bar{\zeta}}Y_{1'} ={}& -(\tfrac{1}{2}h_{nn} + i\psi).
 \label{RelationYmetric3}
\end{align}
\end{subequations}
From these equations we see that there is some freedom 
in $Y_{0'}$, $Y_{1'}$: one can check that the equations 
are invariant under $Y_{0'} \to Y_{0'} + \tau_{0'}$, 
$Y_{1'} \to Y_{1'} + \tau_{1'}$, where 
$\tau_{0'}=p(v,\zeta)\bar{\zeta}+q_{0'}(v,\zeta)$ and 
$\tau_{1'}=p(v,\zeta)u+q_{1'}(v,\zeta)$, 
with $p,q_{0'},q_{1'}$ arbitrary functions of $v,\zeta$.
Alternatively, this is seen from integrating equations \eqref{RelationsYmetric}, which gives
\begin{subequations}\label{ComponentsY}
\begin{align}
 Y_{0'} ={}& \int{\rm d}uh_{\bar{m}\bar{m}} + 
 p(v,\zeta)\bar{\zeta}+q_{0'}(v,\zeta), \label{Y0} \\
 Y_{1'} ={}& -\int{\rm d}\bar{\zeta}(\tfrac{1}{2}h_{nn}+i\psi) 
 +p(v,\zeta)u+q_{1'}(v,\zeta). \label{Y1}
\end{align}
\end{subequations}
From a coordinate-free perspective, the freedom in 
$p,q_{0'},q_{1'}$ corresponds to the fact that 
eq. \eqref{Y} is invariant under $Y_{A'}\to Y_{A'}+\tau_{A'}$, 
where $\tau_{A'}$ is any solution to $\tn_{(A'}\tau_{B')}=0$.
We will not need to use this freedom.

\medskip
In view of \eqref{Y}, the original real metric perturbation is  
\begin{equation}
 h_{ab} = \gamma_{ab} + \bar{\gamma}_{ab}, \qquad
 \gamma_{ab} = o_{A}o_{B}\tn_{(A'}Y_{B')}. \label{hY}
\end{equation}
The linearized Ricci operator for tensor fields of the form 
$\gamma_{ab} = o_{A}o_{B}\tn_{(A'}Y_{B')}$ is, of course, 
a special case of \eqref{RTooX}. 
After some calculations, we find 
\begin{equation}
 -2\dot{R}_{ABA'B'}[\gamma] = 
 o_{A}o_{B}\left[ 2\tn_{A'}\tn_{B'}\nabla^{C'}Y_{C'} 
 -\nabla_{(A'}\tn_{B')}\tn^{C'}Y_{C'} \right] 
 -o_{(A}\iota_{B)}\tn_{A'}\tn_{B'}\tn^{C'}Y_{C'}.
 \label{RToonY}
\end{equation}
Summarizing, so far we have only imposed the vanishing 
of the perturbed Ricci scalar, and we used this to deduce 
the structure \eqref{hY} of the metric perturbation.
Using \eqref{RToonY} and its complex conjugate,
the rest of the Einstein equations \eqref{LinearizedEinsteinEq} is 
\begin{align}
o_{A}o_{B}\left[ 2\tn_{A'}\tn_{B'}\nabla^{C'}Y_{C'} 
 -\nabla_{(A'}\tn_{B')}\tn^{C'}Y_{C'} \right] 
 -o_{(A}\iota_{B)}\tn_{A'}\tn_{B'}\tn^{C'}Y_{C'} 
 +\rm{c.c}
 = 0.
 \label{LEEY}
\end{align}
This equation is automatically satisfied {\em if} $\nabla^{C'}Y_{C'}$ 
and $\tn^{C'}Y_{C'}$ vanish. In this case the result 
of theorem \ref{Thm:Main} would follow immediately: 
the equation $\tn^{C'}Y_{C'}=0$ would imply that 
$Y_{C'}=\tn_{C'}\Phi$ for some (locally defined) complex 
scalar field $\Phi$, and $\nabla^{C'}Y_{C'}=0$ 
would give the wave equation $\Box\Phi=0$.
However, the converse of the above statement is not 
necessarily true: the equation $\dot{R}_{ABA'B'}[h]=0$ does not  
imply that $\nabla^{C'}Y_{C'}$ and $\tn^{C'}Y_{C'}$ vanish. 

\smallskip
The non-vanishing of $\nabla^{C'}Y_{C'}$ and $\tn^{C'}Y_{C'}$ 
makes the completion of the proof of theorem \ref{Thm:Main} 
more difficult.
What we will show is that these fields can be set 
to zero by a gauge transformation.
To this end, recall that we mentioned in section 
\ref{Sec:RadiationGauge} that we still have the freedom to 
perform residual gauge transformations, i.e. transformations 
of the form
\begin{equation}
 h_{ab} \to h'_{ab} = h_{ab} - K[\xi]_{ab}
 \label{GaugeTransformation1}
\end{equation}
where $K[\xi]_{ab}=\nabla_{a}\xi_{b}+\nabla_{b}\xi_{a}$ satisfies 
$\ell^{a}K[\xi]_{ab}=0=g^{ab}K[\xi]_{ab}$.
We analyse this residual freedom in appendix \ref{App:ResidualGauge}, where we show that there exists 
a spinor field $g_{A'}$ such that $K[\xi]_{ab}$ can be 
written as in eq. \eqref{KRG2}. The gauge-transformed metric 
is then
\begin{equation}
 h'_{ab} = \gamma'_{ab} + \bar{\gamma}'_{ab}, 
 \label{h'}
\end{equation}
where 
\begin{equation}
 \gamma'_{ab} = o_{A}o_{B}\tn_{(A'}Z_{B')}, \qquad 
 Z_{B'}=Y_{B'}-g_{B'}.
\end{equation}
Since \eqref{GaugeTransformation1}-\eqref{h'} 
is a gauge transformation, we have 
$\dot{R}_{ABA'B'}[h]=\dot{R}_{ABA'B'}[h']$, thus, 
the Einstein equations \eqref{LEEY} are equivalently
\begin{align}
o_{A}o_{B}\left[ 2\tn_{A'}\tn_{B'}\nabla^{C'}Z_{C'} 
 -\nabla_{(A'}\tn_{B')}\tn^{C'}Z_{C'} \right] 
 -o_{(A}\iota_{B)}\tn_{A'}\tn_{B'}\tn^{C'}Z_{C'} 
 +\rm{c.c}
 = 0.
\end{align}

\begin{proposition}\label{Prop:neutrino}
The gauge transformation \eqref{GaugeTransformation1}-\eqref{h'}
can be chosen such that the spinor field $Z_{A'}$
satisfies the neutrino equation
\begin{equation}
 \nabla^{AA'}Z_{A'} = 0. \label{neutrino}
\end{equation}
\end{proposition}

We defer the proof of this proposition to appendix 
\ref{App:neutrino}.
Now, any solution to \eqref{neutrino} 
can be written (locally) as $Z_{A'}=\tn_{A'}\Phi$ for some 
complex scalar field that satisfies the wave equation.
To see this, first contract \eqref{neutrino} with $o_{A}$, 
which gives $\tn^{A'}Z_{A'}=0$. 
This implies that there is, locally, a complex 
scalar field $\Phi$ such that $Z_{A'}=\tn_{A'}\Phi$ 
(see Remark \ref{Remark:Potentials}).
Contracting now \eqref{neutrino} with $\iota_{A}$, 
we get $\nabla^{A'}\tn_{A'}\Phi=0$, which is the same 
as $\Box\Phi=0$. 

Summarizing, the original real metric perturbation
is $h_{ab}=h'_{ab} + K[\xi]_{ab}$, where 
$h'_{ab}=\gamma'_{ab}+\bar\gamma'_{ab}$, 
$\gamma'_{ab}$ is given by
\begin{equation}
 \gamma'_{ab} = o_{A}o_{B}\tn_{A'}\tn_{B'}\Phi,
\end{equation}
and $\Phi$ satisfies the wave equation $\Box\Phi=0$ 
on the background pp-wave space-time.
This concludes the proof of theorem \ref{Thm:Main}.

\begin{remark}
\begin{enumerate}
\item The perturbation 
$h'_{ab}=\gamma'_{ab}+\bar{\gamma}'_{ab}$ is both in 
radiation gauge and in Lorenz gauge: 
one can check that $\ell^{a}h'_{ab}=0=g^{ab}h'_{ab}$ 
as well as $\nabla^{a}h'_{ab}=0$.
\item The residual radiation gauge freedom is essential 
for the proof of theorem \ref{Thm:Main}. Note that this must also 
be used if one wants to apply the same method to even 
the simplest case of Minkowski space-time, 
which can be obtained by simply setting $H=0$ in our 
formulas\footnote{
Note that the formulations of the Minkowski problem in 
\cite{TorresdelCastillo}, \cite{Penrose65}, \cite[Section 5.7]{PR1}
are very different to our method and only apply when the 
background is flat.}.
\item As a by-product of the above construction, 
we obtained a method to generate solutions to the linearized 
Einstein vacuum equations from solutions to the neutrino equation.
\item If the potential $\Phi$ is independent of $u$, 
i.e. $\Phi_{u}=0$, then the perturbation actually gives a 
solution to the full (non-linear) Einstein equations. 
This can be seen from eqs. \eqref{Hertz}, \eqref{WaveEquation}, 
\eqref{Riccippwave}, \eqref{Boxppwave}: 
$\Phi$ is just a perturbation to the background wave profile $H$.
\end{enumerate}
\end{remark}

\section{Generalization to a ``half-K\"ahler'' vacuum space-time}
\label{Sec:halfkahler}

In this section we briefly show how the ideas of the 
previous sections can be carried over to a more general 
(real, Lorentzian) space-time: the ``half-K\"ahler'' case of 
table \ref{table}.
This is defined by the condition that there is a parallel 
{\em projective} spinor, that is, a `spinor field up to scale'
that is parallel.
We are not aware of a description of this space-time (or its 
complex generalization) analogous to the one given in 
propositions \ref{Prop:ppwave}, \ref{Prop:Complexppwave}. 
In the Euclidean case, the manifold must be K\"ahler, 
so the Lorentzian version might be of interest on its own 
right\footnote{Perhaps a closer Lorentzian analogue to a 
K\"ahler manifold would be a space-time where we have 
both $\Theta_{AA'}o^{B}=0$ and $\Theta_{AA'}\iota^{B}=0$. 
This may be seen as a complexified version of K\"ahler geometry. 
Black holes (in particular) are conformal to this space-time.}. 
In addition, a {\em real} version of the complex result of 
Prop. \ref{Prop:Complexppwave} for this case 
would correspond again to a split signature metric.
Here we restrict ourselves merely to the description 
of gravitational perturbations.

\smallskip
A convenient way of expressing the existence of a spinor 
up to scale that is parallel is to use GHP language 
(cf. \cite[Section 4.12]{PR1}).
As is known, the use of spinors/vectors up to scale in relativity 
brings about the notion of ``GHP weight''.
Let $\{o^{A},\iota^{A}\}$ be two spinor fields in a Lorentzian 
space-time $(M,g_{ab})$, with $o_{A}\iota^{A}=1$. 
A (scalar/tensor/spinor) 
field $\eta$ is said to have GHP weight $\{p,q\}$ if, 
under the rescaling $o^{A}\to\lambda o^{A}$, 
$\iota^{A}\to\lambda^{-1}\iota^{A}$ (with $\lambda$ a complex 
scalar different from zero),
it transforms as $\eta \to \lambda^{p}\bar{\lambda}^{q}\eta$.
A derivative operator that is covariant under this transformation 
is the GHP connection
$\Theta_{a}=\nabla_{a}+p\omega_{a}+q\bar{\omega}_{a}$, 
where $\omega_{a}:=\iota_{B}\nabla_{a}o^{B}$. 
The existence of a parallel projective spinor can 
then be expressed as the condition $\Theta_{AA'}o^{B}=0$.

\smallskip
If $o^{A}$ satisfies $\Theta_{AA'}o^{B}=0$, then a few 
calculations using $[\Theta_{a},\Theta_{b}]o^{C}=0$ show that 
the space-time must be of Petrov type II. 
If in addition, we impose 
the vacuum condition $\Phi_{ABA'B'}=0=\Lambda$,
then $\Psi_{2}=0$, $\nabla_{A(A'}\omega^{A}_{B')}=0$ and 
$\nabla_{A'(A}\omega^{A'}_{B)}=-\Psi_{3}o_{A}o_{B}$.
The GHP connection is then self-dual and algebraically special.
Notice that $\Psi_{2}=0$ excludes the type D case, so in particular, 
black holes are not included in this vacuum class.

\smallskip
The vector field $\ell^{b}=o^{B}\bar{o}^{B'}$ satisfies 
$\Theta_{a}\ell^{b}=0$, so it is tangent to a null congruence 
that is both geodesic and shear-free.
The radiation gauge for gravitational perturbations 
can then be imposed \cite{Whiting}. 
The discussion from now on is analogous to what we did in 
sections \ref{Sec:RadiationGauge} and \ref{Sec:Potentials}, 
the only extra point to keep in mind is that all fields now carry 
GHP weights.
If $h_{ab}$ is a perturbation in radiation gauge, we can 
write it as in \eqref{ooX0}, that is 
$h_{ab}=\gamma_{ab}+\bar\gamma_{ab}$, with 
$\gamma_{ab}=o_{A}o_{B}X_{A'B'}$. 
The spinor field $X_{A'B'}$ has weights $\{-2,0\}$.
The linearized Einstein equations are then 
$\dot{R}_{ab}[h]=
\dot{R}_{ab}[\gamma]+\dot{R}_{ab}[\bar\gamma]=0$, 
and a calculation shows that 
\begin{equation}
 -2\dot{R}_{ab}[\gamma] = 
 o_{A}o_{B}[\Box^{\T}X_{A'B'}+2\tT^{C'}\T_{(A'}X_{B')C'}]
 -2o_{(A}\iota_{B)}\tT^{C'}\tT_{(A'}X_{B')C'}
 -\tfrac{1}{2}g_{ab}\dot{R}[\gamma],
\end{equation}
where 
\begin{equation}
 \dot{R}[\gamma]=\tT^{A'}\tT^{B'}X_{A'B'},
\end{equation}
and we defined 
\begin{equation}
 \Box^{\T} := g^{ab}\T_{a}\T_{b}, \qquad
 \tT_{A'}:=o^{A}\Theta_{AA'}, \qquad \T_{A'}:=\iota^{A}\Theta_{AA'}.
\end{equation}
Choosing the free scalar field $\psi$ in \eqref{ooX0} so 
that $\tT^{A'}\tT^{B'}X_{A'B'}=0$, 
we deduce that there is a spinor field $Y_{A'}$, with 
weights $\{-3,0\}$, such that $X_{A'B'}=\tT_{(A'}Y_{B')}$.
Thus, after some calculations, the linearized Einstein equations 
become (compare to \eqref{LEEY})
\begin{align}
o_{A}o_{B}\left[ 2\tT_{A'}\tT_{B'}\T^{C'}Y_{C'} 
 -\T_{(A'}\tT_{B')}\tT^{C'}Y_{C'} \right] 
 -o_{(A}\iota_{B)}\tT_{A'}\tT_{B'}\tT^{C'}Y_{C'} 
 +\rm{c.c}
 = 0.
 \label{LEEYhk}
\end{align}
For a residual radiation gauge transformation
$h_{ab}\to h'_{ab}=h_{ab}-K[\xi]_{ab}$, 
(where $K[\xi]_{ab}=\nabla_{a}\xi_{b}+\nabla_{b}\xi_{a}$ 
satisfies \eqref{RGresidual})
there is a spinor $g_{A'}$ such that 
$K[\xi]_{ab} = o_{A}o_{B}\tT_{(A'}g_{B')} + {\rm c.c}$. 
So the Einstein equations for $h'_{ab}$ 
are the same as \eqref{LEEYhk} with $Y_{A'}$ replaced by 
$Z_{A'}=Y_{A'}-g_{A'}$.
The analysis of the residual gauge is analogous to the pp-wave 
case discussed in appendix \ref{App:ResidualGauge}, 
where instead of coordinate derivatives we use GHP operators. 
For example, instead of \eqref{SecondDerivativesGV}, we find 
$\tho\xi_{\ell}=0$, 
$\tho^{2}\xi_{m}=\tho^{2}\xi_{\bar{m}}=\tho^{2}\xi_{n}=0$, 
$\edt\edt'\xi_{\ell}=0$.
Choosing then the gauge such that $\tT^{A'}g_{A'}=\tT^{A'}Y_{A'}$, 
$\T^{A'}g_{A'}=\T^{A'}Y_{A'}$, we obtain 
$\tT^{A'}Z_{A'}=0$ and $\T^{A'}Z_{A'}=0$, or equivalently 
a weighted neutrino equation
\begin{equation}
 \T^{AA'}Z_{A'} = 0.
\end{equation}
From $\tT^{A'}Z_{A'}=0$ we deduce that there is, locally, 
a complex scalar field $\Phi$, with weights $\{-4,0\}$, such 
that $Z_{A'}=\tT_{A'}\Phi$, and from $\T^{A'}Z_{A'}=0$ 
we deduce that $\Box^{\Theta}\Phi=0$.

\smallskip
In summary, we see that any real gravitational perturbation 
to a ``half-K\"ahler'' vacuum space-time, once written 
in radiation gauge and assuming that it satisfies the 
linearized Einstein vacuum equations, can be locally expressed as
\begin{equation}
 h_{ab} = o_{A}o_{B}\tT_{A'}\tT_{B'}\Phi + {\rm c.c} 
 + 2\nabla_{(a}\xi_{b)},
\end{equation}
where $\Phi$ satisfies the GHP weighted wave equation 
\begin{equation}
 \Box^{\Theta}\Phi = 0.
\end{equation}
This generalizes the pp-wave result, theorem \ref{Thm:Main}.
Notice that, analogously to the pp-wave case, the Hertz potential 
$h'_{ab}= o_{A}o_{B}\tT_{A'}\tT_{B'}\Phi + {\rm c.c} $ is 
both in radiation gauge and in Lorenz gauge 
(i.e. $\nabla^{a}h'_{ab}=0$).

\section{Conclusions}
\label{Sec:Conclusions}

In this work we have shown that any real, linear gravitational 
perturbation of a (real, Lorentzian) vacuum pp-wave space-time 
can be locally expressed, modulo gauge transformations, 
as the real part of a Hertz/Debye potential, where the 
scalar Debye potential satisfies the wave equation 
of the background pp-wave solution. 
This is believed to hold for more general backgrounds 
as well (replacing the wave equation by, e.g., the 
Teukolsky equation), 
but to our knowledge, the result has been completely proven
only for perturbations of Minkowski 
\cite{TorresdelCastillo}, \cite{Penrose65}, \cite[Section 5.7]{PR1}. 
We stress that our result is local, cf. Remark \ref{Remark:Potentials} 
and also \cite{vanDeMeent}.

\smallskip
We also showed the connections between the Hertz/Debye 
representation for perturbations of pp-waves 
and the non-linear structure of a complex space-time 
with a parallel spinor. 
This illustrates the formal relation between 
this representation and a 
particular case of the hyper-heavenly construction of 
Pleba\'nski and Robinson \cite{PlebanskiRobinson}. 
In addition, we argued that a {\em linear} problem in a real 
space with a Lorentzian metric is related to a {\em non-linear} 
problem also in a real space but with a split signature metric. 
This is interesting in view of modern developments 
where physics in split signature is relevant,
especially in the context of scattering amplitudes and 
connections to gravitation, see \cite{Crawley}, \cite{Kosower}.

\smallskip
Our approach relied on using special 
complex 2-surfaces in the complexified space-time, called 
$\alpha$- and $\beta$-surfaces, which are the basic object 
of twistor theory, cf. \cite{PR2}. 
These surfaces are present, in particular, 
for (complexifications of) any algebraically special, 
vacuum, real, Lorentzian space-time. 
Thus, the method employed in this work can also be applied to the
analysis of linearized gravity on more general backgrounds.
We illustrated this by generalizing our result to perturbations 
of a ``half-K\"ahler'' vacuum space-time.
Explicit computations in more general backgrounds 
are more involved due to the complicated structure 
of the curvature.
The interpretation of the coordinates defined by twistor surfaces 
is also more difficult than in the pp-wave case
(where these coordinates are simply Brinkmann coordinates).

\smallskip
From a physical point of view, our motivation came from
perturbation theory in general relativity and its applications 
to gravitational wave physics, concerning the Hertz/Debye 
potential representation of perturbations, and gauge issues. 
While the currently most interesting space-times
for gravitational wave physics are more general than 
pp-waves, representing e.g. single or binary black holes,
the case of pp-waves already presents conceptual 
difficulties similar to those that appear in the other 
more general cases. This can be seen from our study of 
perturbations to ``half-K\"ahler'' space-times.
The application of these ideas to the general class of 
Petrov type II vacuum solutions (including type D and the 
Kerr solution) is left for future work \cite{AAAW2}.

\smallskip
From a geometric perspective, our motivation 
originated in the relations that (generalized) parallel spinors 
have with complex geometry, 
as discussed in section \ref{Sec:Motivation}.
We focused on perturbations to the simplest case of 
a parallel spinor in Lorentz signature, a pp-wave metric. 
In Euclidean signature this corresponds to
hyper-K\"ahler manifolds, and linear perturbations 
in this context have been studied e.g. in 
\cite{DunajskiMason}, \cite{Alexandrov}. 
Natural generalizations are the other cases described in 
Table \ref{table}. We also studied the Lorentzian
``half-K\"ahler'' case, which in a Riemannian setting 
would correspond to perturbations of K\"ahler manifolds. 
A natural next step would be the study of perturbations to the 
``half-Hermitian'' case $\mathcal{C}_{AA'}o^{B}=0$.

\subsubsection*{Acknowledgements}

I am very grateful to Steffen Aksteiner, Lars Andersson 
and Bernard Whiting for many discussions on the topics 
of this work, and for ongoing collaboration on generalizations of 
the results presented in this manuscript.
The author is supported by the Alexander von Humboldt 
Foundation.

\appendix

\section{Gauge issues}
\label{Appendix:Gauge}

Throughout this appendix we assume a (real, Lorentzian) 
vacuum pp-wave background, with the 
special spin frame $\{o^{A},\iota^{A}\}$,
and its complex conjugate $\{\bar{o}^{A'},\bar{\iota}^{A'}\}$,
introduced in section \ref{Sec:ppwaves}.
The associated null tetrad $\{\ell^{a},n^{a},m^{a},\bar{m}^{a}\}$ 
is defined as in eq. \eqref{nulltetrad}. 
In terms of Brinkmann coordinates (see eq. \eqref{ppwavemetric}), 
we have
\begin{equation}
 \ell^{a}\partial_{a} = \partial_{u}, \qquad 
 m^{a}\partial_{a} = -\partial_{\bar{\zeta}}, \qquad 
 \bar{m}^{a}\partial_{a} = -\partial_{\zeta}, \qquad 
 n^{a}\partial_{a} = \partial_{v} - \tfrac{1}{2}H\partial_{u}. 
 \label{NullTetradCoordinates}
\end{equation}
The connection coefficients are given by
\begin{equation}
 \nabla_{a}\ell^{b} = 0, \qquad 
 \nabla_{a}m^{b} = -\bar{\kappa}'\ell_{a}\ell^{b}, \qquad 
 \nabla_{a}\bar{m}^{b} = -\kappa'\ell_{a}\ell^{b}, \qquad 
 \nabla_{a}n^{b} = 
 -\ell_{a}(\kappa' m^{b}+\bar{\kappa}'\bar{m}^{b}).
 \label{DerNullTetrad}
\end{equation}

\subsection{The gauge operator}\label{App:gauge}

For an arbitrary covector field $\xi_{a}$, we define the ``gauge 
operator'' (or Killing operator) $K$ by
\begin{equation}
 K[\xi]_{ab} = \nabla_{a}\xi_{b} + \nabla_{b}\xi_{a}.
 \label{GaugeOperator}
\end{equation}
In terms of a null tetrad, this can be written as follows:
\begin{align*}
 K[\xi]_{ab} ={}& K_{nn}\ell_{a}\ell_{b} -2K_{n\bar{m}}\ell_{(a}m_{b)}
 -2K_{nm}\ell_{(a}\bar{m}_{b)} +K_{\bar{m}\bar{m}}m_{a}m_{b} 
 +K_{mm}\bar{m}_{a}\bar{m}_{b} \\
 & + K_{\ell\ell} n_{a}n_{b} -2K_{\ell\bar{m}}n_{(a}m_{b)} 
 -2K_{\ell m}n_{(a}\bar{m}_{b)} +2K_{\ell n} n_{(a}\ell_{b)} 
 -2K_{m\bar{m}}m_{(a}\bar{m}_{b)}.
\end{align*}
For a pp-wave, using 
\eqref{NullTetradCoordinates}-\eqref{DerNullTetrad} we find:
\begin{subequations}\label{GOcomponents}
\begin{align}
 & K_{\ell\ell} = 2\partial_{u}\xi_{\ell}, 
 &
 &K_{m\bar{m}} =
 -(\partial_{\bar{\zeta}}\xi_{\bar{m}}+\partial_{\zeta}\xi_{m})
 \\
 &K_{\ell\bar{m}} = \partial_{u}\xi_{\bar{m}} 
 - \partial_{\zeta}\xi_{\ell}, 
 &
 & K_{nm} =  (\partial_{v}-\tfrac{1}{2}H\partial_{u})\xi_{m} 
 +\bar{\kappa}'\xi_{\ell} - \partial_{\bar{\zeta}}\xi_{n}, 
 \\
 &K_{\ell m} = \partial_{u}\xi_{m} - \partial_{\bar{\zeta}}\xi_{\ell}, 
 &
 & K_{n\bar{m}} =(\partial_{v}-\tfrac{1}{2}H\partial_{u})\xi_{\bar{m}} 
 +\kappa'\xi_{\ell} - \partial_{\zeta}\xi_{n}, 
 \\
 & K_{\bar{m}\bar{m}} = -2\partial_{\zeta}\xi_{\bar{m}}, 
 &
 & K_{\ell n} = 
 \partial_{u}\xi_{n} + (\partial_{v}-\tfrac{1}{2}H\partial_{u})\xi_{\ell}, 
 \\
 & K_{mm} = -2\partial_{\bar{\zeta}}\xi_{m}, 
 &
 & K_{nn} = 2(\partial_{v}-\tfrac{1}{2}H\partial_{u})\xi_{n} 
 +2\kappa'\xi_{m} + 2\bar{\kappa}'\xi_{\bar{m}}.
\end{align}
\end{subequations}

\subsubsection{Residual radiation gauge freedom}
\label{App:ResidualGauge}

The radiation gauge \eqref{RadiationGauge} is preserved by 
transformations in which the new gauge vector $\xi_{a}$ satisfies
\begin{equation}
 \ell^{a}K[\xi]_{ab} = 0, \qquad g^{ab}K[\xi]_{ab} = 0.
 \label{RGresidual}
\end{equation}
Equivalently: $K_{\ell\ell}=K_{\ell m}=K_{\ell \bar{m}}=K_{\ell n} 
=K_{m\bar{m}}=0$. 
Using identities \eqref{GOcomponents}, this is
\begin{subequations}\label{RGresidualComponents}
\begin{align}
 \partial_{u}\xi_{\ell} ={}& 0, \\
 \partial_{u}\xi_{\bar{m}} - \partial_{\zeta}\xi_{\ell} ={}& 0, \\
 \partial_{u}\xi_{m} - \partial_{\bar{\zeta}}\xi_{\ell} ={}& 0, \\
 \partial_{u}\xi_{n} + \partial_{v}\xi_{\ell} ={}& 0, \\
 \partial_{\bar{\zeta}}\xi_{\bar{m}} + \partial_{\zeta}\xi_{m} ={}& 0.
\end{align}
\end{subequations}
From here we deduce 
\begin{equation}
 \partial^{2}_{u}\xi_{\bar{m}} = \partial^{2}_{u}\xi_{m} 
 = \partial^{2}_{u}\xi_{n} = 0, \qquad 
 \partial_{\zeta}\partial_{\bar{\zeta}}\xi_{\ell} = 0. 
 \label{SecondDerivativesGV}
\end{equation}
Notice that in view of \eqref{GOcomponents}, 
and given that $\partial_{u}$ is a Killing vector of the background 
space-time, from \eqref{SecondDerivativesGV} it follows that
$\partial^{2}_{u}K_{\bf ab}=0$ for all ${\bf a,b}=u,v,\zeta,\bar{\zeta}$. 

So we get the following general form for the components 
of the (real) gauge vector $\xi_{a}$:
\begin{subequations}\label{Componentsgaugevector}
\begin{align}
 \xi_{\ell} ={}& f_{1}(v,\zeta) + f_{2}(v,\bar{\zeta}), \\
 \xi_{\bar{m}} ={}& [\partial_{\zeta}f_{1}(v,\zeta)] u 
 + f_{3}(v,\zeta,\bar{\zeta}),\\
 \xi_{n} ={}& -[\partial_{v}f_{1}(v,\zeta)
 +\partial_{v}f_{2}(v,\bar{\zeta})]u + f_{4}(v,\zeta,\bar{\zeta}),
\end{align}
\end{subequations}
for some functions $f_{1}(v,\zeta)$, $f_{2}(v,\bar\zeta)$, 
$f_{3}(v,\zeta,\bar{\zeta})$ and $f_{4}(v,\zeta,\bar{\zeta})$. 
Apart from reality conditions for $\xi_{a}$,
any restrictions on these functions will be differential.

Given that we here impose the gauge operator 
\eqref{GaugeOperator} to satisfy \eqref{RGresidual}, 
the same reasoning that we used in section \ref{Sec:RadiationGauge} to deduce \eqref{ooX0}
now gives
\begin{equation}
 K[\xi]_{ab} = 
 o_{A}o_{B}G_{A'B'} + \bar{o}_{A'}\bar{o}_{B'}\bar{G}_{AB},
 \label{KRG}
\end{equation}
where 
\begin{equation}
 G_{A'B'}=(\tfrac{1}{2}K_{nn}+i\eta)\bar{o}_{A'}\bar{o}_{B'} 
 -2K_{n\bar{m}}\bar{o}_{(A'}\bar{\iota}_{B')}
 +K_{\bar{m}\bar{m}}\bar{\iota}_{A'}\bar{\iota}_{B'}
 \label{SpinorForGauge}
\end{equation}
and we have included an arbitrary scalar field $\eta$. 
Explicit expressions for $K_{nn}$, $K_{n\bar{m}}$, 
$K_{\bar{m}\bar{m}}$ in terms of $\xi_{a}$ are given in 
\eqref{GOcomponents}.
We can now express $G_{A'B'}$ in terms of a 1-index spinor 
by doing the same trick that we did in section \ref{Sec:Potentials}.
The linearized Ricci scalar of \eqref{KRG} is
$ \dot{R}[K[\xi]] = \tn^{A'}\tn^{B'}G_{A'B'} 
 +\bn^{A}\bn^{B}\bar{G}_{AB}$ 
(which vanishes identically since the background is vacuum), 
where
\begin{equation*}
  \tn^{A'}\tn^{B'}G_{A'B'}  = 
  \partial^{2}_{\bar{\zeta}}K_{\bar{m}\bar{m}} 
 +2\partial_{u}\partial_{\bar{\zeta}} K_{n\bar{m}}
 +\tfrac{1}{2}\partial^{2}_{u}K_{nn}+i\partial^{2}_{u}\eta.
\end{equation*}
A short calculation shows that 
$\partial^{2}_{u}K_{nn}=0
=\partial_{u}\partial_{\bar{\zeta}} K_{n\bar{m}}$, 
while $\partial^{2}_{\bar{\zeta}}K_{\bar{m}\bar{m}}$ is independent 
of $u$. Therefore, if we choose the arbitrary scalar $\eta$ 
in \eqref{SpinorForGauge} in the form
$\eta=
\tfrac{i}{2}[\partial^{2}_{\bar{\zeta}}K_{\bar{m}\bar{m}}]u^{2}$, 
then $\tn^{A'}\tn^{B'}G_{A'B'}=0$, so there is a 
spinor field $g_{A'}$ such that $G_{A'B'} = \tn_{(A'}g_{B')}$, 
and 
\begin{equation}
K[\xi]_{ab} = o_{A}o_{B}\tn_{(A'}g_{B')} 
 + \bar{o}_{A'}\bar{o}_{B'}\bn_{(A}\bar{g}_{B)}.
 \label{KRG2}
\end{equation}
The relation between $g_{A'}$ and $\xi_{a}$ is given by 
\begin{subequations}\label{dergaugespinor}
\begin{align}
 \partial_{u}g_{0'} ={}& K_{\bar{m}\bar{m}}, \\
 \partial_{u}g_{1'} - \partial_{\bar{\zeta}}g_{0'} ={}& 2K_{n\bar{m}}, \\
 \partial_{\bar{\zeta}}g_{1'} ={}& -(\tfrac{1}{2}K_{nn}+i\eta),
\end{align}
\end{subequations}
where in the right-hand sides one replaces the expressions  
\eqref{GOcomponents}.
We deduce from here that 
\begin{equation}
 \partial^{3}_{u}g_{0'} = 0, \qquad 
 \partial_{\bar{\zeta}}\partial^{2}_{u}g_{0'} = 0, \qquad 
 \partial^{3}_{u}g_{1'} = 0. 
 \label{Restrictionsg}
\end{equation}
Thus, the general structure of $g_{0'}$, $g_{1'}$ is 
\begin{subequations}\label{Componentsg}
\begin{align}
 g_{0'} ={}& 
 A_{0'}(v,\zeta)u^{2} + B_{0'}(v,\zeta,\bar{\zeta})u 
 +C_{0'}(v,\zeta,\bar{\zeta}), \\
 g_{1'} ={}& 
  A_{1'}(v,\zeta,\bar\zeta)u^{2} + B_{1'}(v,\zeta,\bar{\zeta})u 
 +C_{1'}(v,\zeta,\bar{\zeta}), 
\end{align}
\end{subequations}
for some functions $A_{0'},...,C_{1'}$ 
where the arguments are as specified in the previous equations.
Using \eqref{dergaugespinor} and \eqref{GOcomponents}, 
one can relate these functions to the ones appearing in 
\eqref{Componentsgaugevector}; this way we see, for example, 
that they do not identically vanish. 
For instance, we get 
$A_{0'}(v,\zeta)=-\partial^{2}_{\zeta}f_{1}(v,\zeta)$.
However,  in general the explicit expressions do not seem 
to be particularly enlightening.

\subsection{Proof of proposition \ref{Prop:neutrino}}
\label{App:neutrino}

We have the identity $\nabla^{AA'}Z_{A'}=
o^{A}\nabla^{A'}Z_{A'}-\iota^{A}\tn^{A'}Z_{A'}$, 
so $\nabla^{AA'}Z_{A'}=0$ iff $\nabla^{A'}Z_{A'}=0$ and 
$\tn^{A'}Z_{A'}=0$. 
Since $Z_{A'}=Y_{A'}-g_{A'}$, we have
\begin{subequations}\label{divZ}
\begin{align}
 \tn^{A'}Z_{A'} ={}& \tn^{A'}Y_{A'} - \tn^{A'}g_{A'}, \\
 \nabla^{A'}Z_{A'} ={}& \nabla^{A'}Y_{A'} - \nabla^{A'}g_{A'},
\end{align}
\end{subequations}
so we want to show that, as long as the Einstein equations are 
satisfied, we can choose the gauge transformation such that 
the associated spinor field $g_{A'}$ satisfies 
$\tn^{A'}g_{A'}=\tn^{A'}Y_{A'}$ and 
$\nabla^{A'}g_{A'}=\nabla^{A'}Y_{A'}$.
 
The first observation is that any requirement for 
the function $\tn^{A'}g_{A'}$ restricts $g_{A'}$ only up to 
the addition of terms of the form $\tn_{A'}S$. 
In other words, we can write $g_{A'}=V_{A'}+2\tn_{A'}S$ 
where $V_{A'}$ and $S$ are independent,
then \eqref{divZ} become
\begin{subequations}\label{divZ2}
\begin{align}
 \tn^{A'}Z_{A'} ={}& \tn^{A'}Y_{A'} - \tn^{A'}V_{A'}, \\
 \nabla^{A'}Z_{A'} ={}& \nabla^{A'}Y_{A'} - \nabla^{A'}V_{A'} -\Box S
\end{align}
\end{subequations}
(where we used the identity $\Box S= 2\nabla^{A'}\tn_{A'}S$),
and we want to show that $V_{A'}$ and $S$ can be chosen 
such that $\tn^{A'}Z_{A'}=0=\nabla^{A'}Z_{A'}$.
Restrictions on $V_{A'}$ and $S$ arise from the fact that 
they come from a gauge transformation: 
the general form of the components of $g_{A'}$ 
was obtained in \eqref{Componentsg}.  
See \eqref{ComponentsVS} below.

\smallskip
In order to obtain expressions for the fields $\tn^{A'}Y_{A'}$, 
$\nabla^{A'}Y_{A'}$, we use the linearized Einstein equations 
\eqref{LEEY}. The non-trivial, independent components are:
\begin{subequations}\label{LinEinsteinComponents}
\begin{align}
 & \partial^{2}_{u}(\tn^{A'}Y_{A'}) = 0, \label{LEEoo1} \\
 & 2\partial^{2}_{u}(\nabla^{A'}Y_{A'}) 
 + \partial_{u}\partial_{\zeta}(\tn^{A'}Y_{A'}) = 0, \label{LEEoo2} \\
 & \partial_{u}\partial_{\bar\zeta}(\tn^{A'}Y_{A'}) 
 +\partial_{u}\partial_{\zeta}(\bn^{A}\bar{Y}_{A}) = 0, 
 \label{LEEoi1} \\
 & 2\partial_{u}\partial_{\bar\zeta}(\nabla^{A'}Y_{A'}) 
 +\tfrac{1}{2}(\partial_{u}\partial_{v}
 +\partial_{\zeta}\partial_{\bar\zeta})(\tn^{A'}Y_{A'}) 
 -\tfrac{1}{2}\partial^{2}_{\zeta}(\bn^{A}\bar{Y}_{A}) = 0, 
 \label{LEEoi2}\\
 & \partial^{2}_{\bar\zeta}(\tn^{A'}Y_{A'}) 
 -4\partial_{u}\partial_{\zeta}(\nabla^{A}\bar{Y}_{A}) 
 -2(\partial_{u}\partial_{v}
 +\partial_{\zeta}\partial_{\bar\zeta})(\bn^{A}\bar{Y}_{A}) = 0, 
 \label{LEEii1}\\
 & 2\partial^{2}_{\bar\zeta}(\nabla^{A'}Y_{A'}) 
 +\partial_{\bar\zeta}(\partial_{v}-\tfrac{1}{2}H\partial_{u})
 (\tn^{A'}Y_{A'}) 
 +2\partial^{2}_{\zeta}(\nabla^{A}\bar{Y}_{A}) 
 +\partial_{\zeta}(\partial_{v}-\tfrac{1}{2}H\partial_{u})
 (\bn^{A}\bar{Y}_{A}) = 0. 
 \label{LEEii2}
\end{align}
\end{subequations}
Equations \eqref{LEEoo1}-\eqref{LEEoo2} correspond to 
$\bar{o}^{A'}\bar{o}^{B'}\dot{R}_{ABA'B'}[h]=0$, 
\eqref{LEEoi1}-\eqref{LEEoi2} correspond to 
$\bar{o}^{A'}\bar{\iota}^{B'}\dot{R}_{ABA'B'}[h]=0$, 
and \eqref{LEEii1}-\eqref{LEEii2} correspond to 
$\bar{\iota}^{A'}\bar{\iota}^{B'}\dot{R}_{ABA'B'}[h]=0$. 
Notice that \eqref{LEEoo1}-\eqref{LEEoo2} are the only 
equations that involve only $\tn^{A'}Y_{A'}$ and
$\nabla^{A'}Y_{A'}$ (not their complex conjugates); 
this is because of the identities \eqref{ooRicci}. 
We will then use \eqref{LEEoo1}-\eqref{LEEoo2} to deduce the 
structure of  $\tn^{A'}Y_{A'}$, $\nabla^{A'}Y_{A'}$.

From \eqref{LEEoo1} we deduce that 
\begin{equation}
 \tn^{A'}Y_{A'} = a(v,\zeta,\bar\zeta) u +b(v,\zeta,\bar\zeta)
 \label{tnY}
\end{equation}
for some functions $a,b$. 
These functions can be written in terms of the metric perturbation 
$h_{ab}$ (up to an arbitrary function of $v,\zeta$, that we can 
set to zero), by noticing that 
$\tn^{A'}Y_{A'}=-(\partial_{u}Y_{1'}+\partial_{\bar\zeta}Y_{0'})$ 
and using eqs. \eqref{Y0}-\eqref{Y1}. 
Taking a $u$-derivative in \eqref{LEEoo2} and using 
\eqref{LEEoo1}, we see that 
$\partial^{3}_{u}(\nabla^{A'}Y_{A'})=0$, so 
$\nabla^{A'}Y_{A'}$ is quadratic in $u$. 
Using also \eqref{tnY}, it follows that 
\begin{equation}
 \nabla^{A'}Y_{A'} = [-\tfrac{1}{4}a_{\zeta}(v,\zeta,\bar\zeta)]u^{2} 
 + c(v,\zeta,\bar\zeta)u + d(v,\zeta,\bar\zeta)
 \label{nY}
\end{equation}
for some functions $c,d$.
The rest of the equations in \eqref{LinEinsteinComponents} 
involve also the complex conjugate fields, and they give 
additional restrictions on the functions that appear in the 
right hand sides of \eqref{tnY}-\eqref{nY}.

Notice that, since the fields $\tn^{A'}g_{A'}$, $\nabla^{A'}g_{A'}$ 
come from a gauge transformation, the same equations 
\eqref{LinEinsteinComponents} hold for them. 
In other words, any restrictions on $\tn^{A'}Y_{A'}$, 
$\nabla^{A'}Y_{A'}$ coming from \eqref{LinEinsteinComponents} 
are also satisfied by $\tn^{A'}g_{A'}$, $\nabla^{A'}g_{A'}$. 
But $\tn^{A'}g_{A'}$, $\nabla^{A'}g_{A'}$ must also fulfil 
restrictions that come from the gauge condition. 
These additional restrictions were analysed in appendix 
\ref{App:ResidualGauge}, where the general expressions 
\eqref{Componentsg} were found. 
In our current context, we have $g_{A'}=V_{A'}+2\tn_{A'}S$, 
or in components $g_{0'}=V_{0'}+2\partial_{u}S$, 
$g_{1'}=V_{1'}-2\partial_{\bar\zeta}S$. 
The restrictions \eqref{Restrictionsg} together 
with the fact that $V_{A'}$ and $S$ are independent 
imply that $\partial^{3}_{u}V_{0'} = 
\partial_{\bar{\zeta}}\partial^{2}_{u}V_{0'} = 
\partial^{3}_{u}V_{1'} = 0$ 
and $\partial^{4}_{u}S=\partial^{3}_{u}\partial_{\bar\zeta}S=0$. 
So we have the following form:
\begin{subequations}\label{ComponentsVS}
\begin{align}
 V_{0'} ={}& 
 \alpha_{0'}(v,\zeta)u^{2} + \beta_{0'}(v,\zeta,\bar{\zeta})u 
 +\gamma_{0'}(v,\zeta,\bar{\zeta}), \\
 V_{1'} ={}& 
  \alpha_{1'}(v,\zeta,\bar\zeta)u^{2} + \beta_{1'}(v,\zeta,\bar{\zeta})u 
 +\gamma_{1'}(v,\zeta,\bar{\zeta}), \\
 S ={}& S_{3}(v,\zeta)u^{3}+S_{2}(v,\zeta,\bar\zeta)u^{2}
 +S_{1}(v,\zeta,\bar\zeta)u + S_{0}(v,\zeta,\bar\zeta).
\end{align}
\end{subequations}
After some tedious calculations, this gives (using in particular 
the expression \eqref{Boxppwave} for $\Box$):
\begin{align}
 \tn^{A'}g_{A'} ={}& -(2\alpha_{1'}+\partial_{\bar\zeta}\beta_{0'})u
 -(\beta_{1'}+\partial_{\bar\zeta}\gamma_{0'}), \\
\nonumber \nabla^{A'}g_{A'} ={}& 
 [\partial_{v}\alpha_{0'}+\partial_{\zeta}\alpha_{1'}
 +6\partial_{v}S_{3}-2\partial_{\zeta}\partial_{\bar\zeta}S_{2}]u^{2}
 \\
\nonumber & +[\partial_{v}\beta_{0'}-H\alpha_{0'}
  +\partial_{\zeta}\beta_{1'}+4\partial_{v}S_{2}
  -2\partial_{\zeta}\partial_{\bar\zeta}S_{1} -6HS_{3}]u \\
 & + [\partial_{v}\gamma_{0'}-\tfrac{1}{2}H\beta_{0'} 
 +\partial_{\zeta}\gamma_{1'}+2\partial_{v}S_{1} 
 -2\partial_{\zeta}\partial_{\bar\zeta}S_{0}-2H S_{2}]
\end{align}
Comparing these expressions to \eqref{tnY}-\eqref{nY}, 
we see that we can choose the free functions in 
\eqref{ComponentsVS} so that $\tn^{A'}g_{A'}=\tn^{A'}Y_{A'}$ 
and $\nabla^{A'}g_{A'}=\nabla^{A'}Y_{A'}$, 
which is what we wanted to prove.

\end{document}